\documentstyle[10pt,emulateapj]{article}

\font\rmsmall=cmr8
\def\smalleps@scaling{1.2}
\def\plotonesmall#1{\centering \leavevmode
    \epsfxsize=\smalleps@scaling\columnwidth \epsfbox{#1}}
\def\etal{{\it et~al\/}.}
\def\eg{{\rm e.g.},}
\def\ie{{\rm i.e.},}

\def\alwaysmath#1{\ifmmode {#1} 
                  \else {$#1\mkern-5mu$} \fi}
\def\kms{\alwaysmath{\,\km\s^{-1}}}

\def\cmtwo{\alwaysmath{\,{\rm cm}^{-2}}}
\def\cmthree{\alwaysmath{\,{\rm cm}^{-3}}}
\def\K{\alwaysmath{\,{\rm K}}}

\def\km{\alwaysmath{\,{\rm  km}}}

\def\s{\alwaysmath{\,{\rm  s}}}
\def\persec{\alwaysmath{\,{\rm  s^{-1}}}}
 
\def\pc{\alwaysmath{\,{\rm  pc}}}

\def\mjysr{\alwaysmath{\,{\rm MJy}\,{\rm sr}^{-1}}}
\def\expo#1{\alwaysmath{10^{#1}}}
\def\nexpo#1{\alwaysmath{\times 10^{#1}}}

\def\hii{H~{\rmsmall II}}
\def\hi{H~{\rmsmall I}}
\def\htwo{\alwaysmath{{\rm H}_2}}
\def\carbon#1{\ifmmode{{^{#1}{\rm C}}}\else{{$^{#1}{\rm C}$}}\fi}
\def\ox#1{\ifmmode{{^{#1}{\rm O}}}\else{{$^{#1}{\rm O}$}}\fi}
\def\cor {\carbon{13}{\rm O}}
\def\co  {\carbon{12}{\rm O}}
\def\tmb{\alwaysmath{{\rm T}_{\litl MB}}}
\def\ta{\alwaysmath{{\rm T}_{\litl A}}}
\def\tex  {\ifmmode{{\rm T_{\rm ex}}}\else{{${\rm T_{\rm ex}}$}}\fi}
\def\tk  {\ifmmode{{\rm T_{\rm k}}}\else{{${\rm T_{\rm k}}$}}\fi}
\def\litl{\rm\scriptscriptstyle}
\def\dtrms{\ifmmode{{\Delta{\rm T_{\litl RMS}}}}\else{{$\Delta{\rm T_{\litl RMS}}$}}\fi}
\def\vlsr{\ifmmode{{\rm V_{\litl LSR}}}\else{{$\rm V_{\litl LSR}$}}\fi}
\def\fwhm{\ifmmode{{\Delta{\rm V}_{\litl FWHM}}}\else{{$\Delta{\rm V}_{\litl FWHM}$}}\fi}
\def\pp{\ifmmode{{^3{\rm P}_{\litl 1}\rightarrow {^3\rm P}_{\litl 0}}}\else{{$
^3{\rm P}_{\litl 1}\rightarrow {^3\rm P}_{\litl 0}$}}\fi}
\def\ppp{\ifmmode{{^2{\rm P}_{\litl 3/2}\rightarrow {^2\rm P}_{\litl 1/2}}}\else{{$
^2{\rm P}_{\litl 3/2}\rightarrow {^2\rm P}_{\litl 1/2}$}}\fi}
\def\ci{{\rm C}~{\rmsmall I}}

\def\av{\alwaysmath{{\rm A}_{\rm v}}}
\def\nnh{\ifmmode{{{\rm N}_{\litl H}}}\else{{${\rm N}_{\litl H}$}}\fi}
\def\nc{\ifmmode{{{\rm N}_{\litl C}}}\else{{${\rm N}_{\litl C}$}}\fi}
\def\ncii{\ifmmode{{{\rm N}_{\litl CII}}}\else{{${\rm N}_{\litl CII}$}}\fi}
\def\nco{\ifmmode{{{\rm N}_{\litl CO}}}\else{{${\rm N}_{\litl CO}$}}\fi}
\def\ncodv{\ifmmode{{{\rm N}_{\litl CO}/{\Delta{\rm V}}}}\else{{${\rm N}_{\litl CO}/{\Delta{\rm V}}$}}\fi}
\def\xco{\ifmmode{{{\rm X}_{\litl CO}}}\else{{${\rm X}_{\litl CO}$}}\fi}
\def\ncrit{\alwaysmath{{\rm n}_{\litl crit}}}
\def\tkin{\alwaysmath{\rm T_{\litl k}}}

\def\irasd{\alwaysmath{{\rm I}_{\litl 100}}}
\def\irasc{\alwaysmath{{\rm I}_{\litl 60}}}

\def\ico{\alwaysmath{{\rm I}_{\litl CO}}}

\begin{document}
\submitted{Accepted for publication in the Astrophysical Journal.  Final
Draft version 1999 November 4.}
\title{Physical State of Molecular Gas in High Galactic Latitude 
Translucent Clouds}
\author{James G. Ingalls\altaffilmark{1} and T.M. Bania}
\affil{Department of Astronomy; Boston University; 725 Commonwealth Avenue,
Boston, MA 02215;\\ ingalls@ipac.caltech.edu, bania@bu.edu}
\and
\author{Adair P. Lane, Matthias Rumitz\altaffilmark{2}, and Antony A. Stark}
\affil{Harvard--Smithsonian Center for Astrophysics; 60 Garden Street, Mail Stop 78,
Cambridge, MA 02138 \\ adair@cfa.harvard.edu, Rumitz@lrz.tu-muenchen.de,
 aas@cfa.harvard.edu}
\altaffiltext{1}{Current address:  Infrared Processing and Analysis Center, Jet
Propulsion Laboratory, California Institute of Technology, 770 South Wilson
Avenue, Pasadena CA 91125}
\altaffiltext{2}{Current address:  Institute for Anatomy, TU--Munich, 
Biedersteiner Str. 29, 80802 Munich, Germany}

\begin{abstract}

The rotational transitions of carbon monoxide
(CO) are the primary means of investigating the density and velocity
structure of the molecular interstellar medium.  Here we study 
the lowest four rotational transitions of CO towards high--latitude
translucent molecular clouds (HLCs).  We report new observations
of the J = (4--3), (2--1), and (1--0) transitions of CO towards
eight high--latitude clouds.  The 
new observations are combined with data from the literature to show that 
the emission from {\it all} observed CO transitions is linearly 
correlated.  This implies
that the excitation conditions which lead to emission in these transitions 
are uniform throughout the clouds.  Observed \cor/\co\  (1--0) integrated 
intensity ratios are generally much greater than the expected abundance 
ratio of the two species, indicating that the regions which emit \co\ 
(1--0) radiation are optically thick.  We develop a statistical method 
to compare the observed line ratios with models of CO excitation and 
radiative transfer.  This enables us to determine the most likely portion 
of the physical parameter space which is compatible with the 
observations.  The model enables us to rule out CO gas temperatures 
greater than $\sim 30\K$,
since the most likely high-temperature configurations are 1\pc -sized 
structures aligned along the line of sight.  The most probable solution 
is a high density and low temperature (HDLT) solution, with 
volume density, $n=\expo{4.5\pm 0.5}\cmthree$, kinetic temperature,
$\tk \approx 8\K$, and CO column density per velocity interval
$\ncodv = \expo{16.6\pm 0.3}\cmtwo/(\kms)$.  The 
CO cell size is $L\sim 0.01\pc$ ($\sim 2000\,$AU).  These cells
are thus tiny fragments within the $\sim$100 times larger CO-emitting
extent of a typical high--latitude cloud.  
We discuss the physical implications of HDLT cells, and we suggest ways 
to test for their existence.
\end{abstract}

\section{Introduction}

The rotational transitions of carbon monoxide (CO) are the most important
probes used to trace the density and velocity structure
of molecular clouds.  Due to the low electric dipole moment of the
CO molecule, the J = (1--0) and (2--1) transitions
are easily excited into emission in molecular clouds with
modest extinctions ($\av\gtrsim 1\,$ magnitude).  Because of this and the
fact that the frequencies of the (1--0) and (2--1) transitions 
lie in two relatively transparent atmospheric windows, they
are the most commonly observed molecular emission lines in the interstellar 
medium (ISM).  Unfortunately 
the information given by observing these transitions is limited, since 
they are not sensitive to volume densities above the critical 
density for collisional excitation, $\ncrit\sim 10^3\cmthree$.
  To probe gas at higher density demands that transitions with higher 
energies be observed.  
Star-formation regions are bright in the high-J CO transitions, but 
modelling these transitions is difficult because (1)
the large CO column density in such regions renders their interiors
completely opaque, and (2) randomly distributed energy sources (stars) 
increase dramatically the complexity of the problem.  Modelling success 
is more likely 
in regions which are relatively optically thin and far from star formation.

The simplest class of molecular cloud observable in CO is the 
translucent cloud (van Dishoeck \& Black 1988).  Translucent clouds
have modest visual extinctions ($\av\approx 1-5\,$mag), and therefore
are intermediate between low-extinction diffuse clouds, which contain 
few molecules observable in emission, and optically thick dark 
clouds.  The high Galactic latitude ($b\gtrsim 15\arcdeg$) molecular clouds
(HLCs; see review by Magnani, Hartmann, \& Speck 1996, and
references therein) are a set of nearby ($\langle d \rangle \approx 
105\pc$; Magnani \etal\ 1996) translucent clouds associated with {\sl IRAS}
cirrus (Low \etal\ 1984).  The HLCs are generally
far from sources of far-ultraviolet (FUV) (6\,eV $< h\nu <$ 13.6\,eV) 
radiation and most are probably exposed to the average interstellar FUV 
radiation field (see van Dishoeck \& Black 1988; Draine 1978; Habing 1968).  
  High--latitude clouds are photochemically and energetically the simplest
molecular objects in the ISM.  They have equal amounts of
carbon gas in atomic form, \ci ,  as in CO, with an average C/CO column 
density ratio of $\sim 1$ (Stark \& van Dishoeck 1994; Ingalls, Bania, \& 
Jackson 1994; Ingalls, \etal\ 1997).  They are essentially isolated 
photodissociation regions (Tielens \& Hollenbach 1985) whose structure is
dominated by the transition from atomic to molecular gas which occurs on 
the surface of all molecular clouds.  Most HLCs are uncomplicated
by internal energy sources.  Some of the HLCs contain T Tauri
stars (Magnani, Caillault, Buchalter, \& Beichman 1995; Pound 1996)
and others have marginally bound dense cores (Reach \etal\ 1995), 
but on the whole HLCs are minimally affected by stars.  Thus they are 
the most basic clouds observable in CO emission and are ideal candidates for 
multi-transition CO studies.

The most comprehensive study of CO in translucent 
and high--latitude clouds was done by van Dishoeck, Black, Phillips, 
\& Gredel (1991).  They observed 4 transitions of \co\ and \cor\ in a 
sample of $\sim 25$ clouds.  They derived relatively low gas volume densities
($n = 200-5000\cmthree$) and high kinetic temperatures ($\tk > 20\K$), but 
they warned that their data are also compatible with higher densities and lower
temperatures.  An ongoing investigation of the physics and 
 chemistry of translucent clouds (Turner 1994; see also Turner, Terzieva, \& 
 Herbst 1999 and references therein) supports the low density, high temperature
conclusion.  On the
 other hand, a study of CO emission towards the translucent edge of a 
cloud in the Perseus--Auriga complex (Falgarone \& Phillips 1996, hereafter 
FP) claimed that the 
CO emission is produced by beam--diluted clumps with high densities 
($n > \expo{4}\cmthree$) and low temperatures ($\tkin\lesssim
15\K$) (see also Falgarone, Phillips, \& Walker 1991).  Interestingly, FP 
admitted that the observations could also be produced by
a more homogeneous CO medium with lower densities 
($n\sim 200\cmthree$) and higher temperatures ($\tk\sim 30-50\K$).  

In general the solution of the statistical equilibrium equations rests on
an assumption about either the kinetic temperature or the cloud
geometry (van Dishoeck \etal\ 1991).  Clearly such solutions are not unique, 
since collisional excitation rates are 
affected by both temperature and density.  Observations
of the rotational transitions of high dipole-moment molecules in 
translucent and high--latitude clouds (Reach, 
Pound, Wilner \& Lee 1995; Heithausen, Corneilussen, \& Grossman 1998) 
support the assertion made by FP that significant quantities of dense gas 
exist, but they do not prove that most of the CO emission originates in such 
gas.  Even the high-density conclusions of Reach \etal\
(1995), and Heithausen \etal\ (1998) rely on assumed low \tk\ values
(5--20\K).  

The CO emission towards translucent clouds
shows some surprising regularities which should be exploited in a 
multitransition study.  For example, 
the \co\ (2--1) intensity is linearly proportional to the (1--0) intensity in 
clouds without star formation, for all positions and velocities observed (FP; 
Falgarone \etal\ 1998).  This important fact implies that the excitation 
conditions which lead to emission in these transitions are uniform
throughout the clouds.  FP 
argued that the simplest situation which would produce uniform excitation
conditions was the high density, low temperature (HDLT) one 
($n> \expo{4}\cmthree;\tk\sim 10\K$), since the
CO level populations would be thermal under these conditions.  They
eliminated the low density, high temperature (LDHT) case 
($n< \expo{4}\cmthree;\tk\gtrsim 20\K$) because the implied 
line-of-sight structure sizes (column density divided by volume density) 
are greater than the $\sim 0.06$\pc\ size of the smallest structure 
observed in projection on the plane of the sky.  Implicit in their argument
is the assumption that the size of a structure along the line-of-sight equals 
(on average) the size of the smallest observed structure in the 
plane of the sky.  

This may not be strictly true.  The best direct evidence for the
existence of tiny CO clumps in translucent clouds is the 
population of unresolved $\sim 2000$\,AU (0.01\pc~)--sized structures seen 
in velocity channel maps of the first {\sl IRAM} key-project 
(Falgarone \etal\ 1998).  These dense unresolved cores are not isolated 
spheres, however, but usually have the shape of elongated filaments
which are unresolved in one direction only.  Thus there is little 
reason to suppose that the average line of 
sight size of CO emission cells equals the smallest observed 
plane-of-sky structure size.

\begin{figure*}[hb!]
\plotfiddle{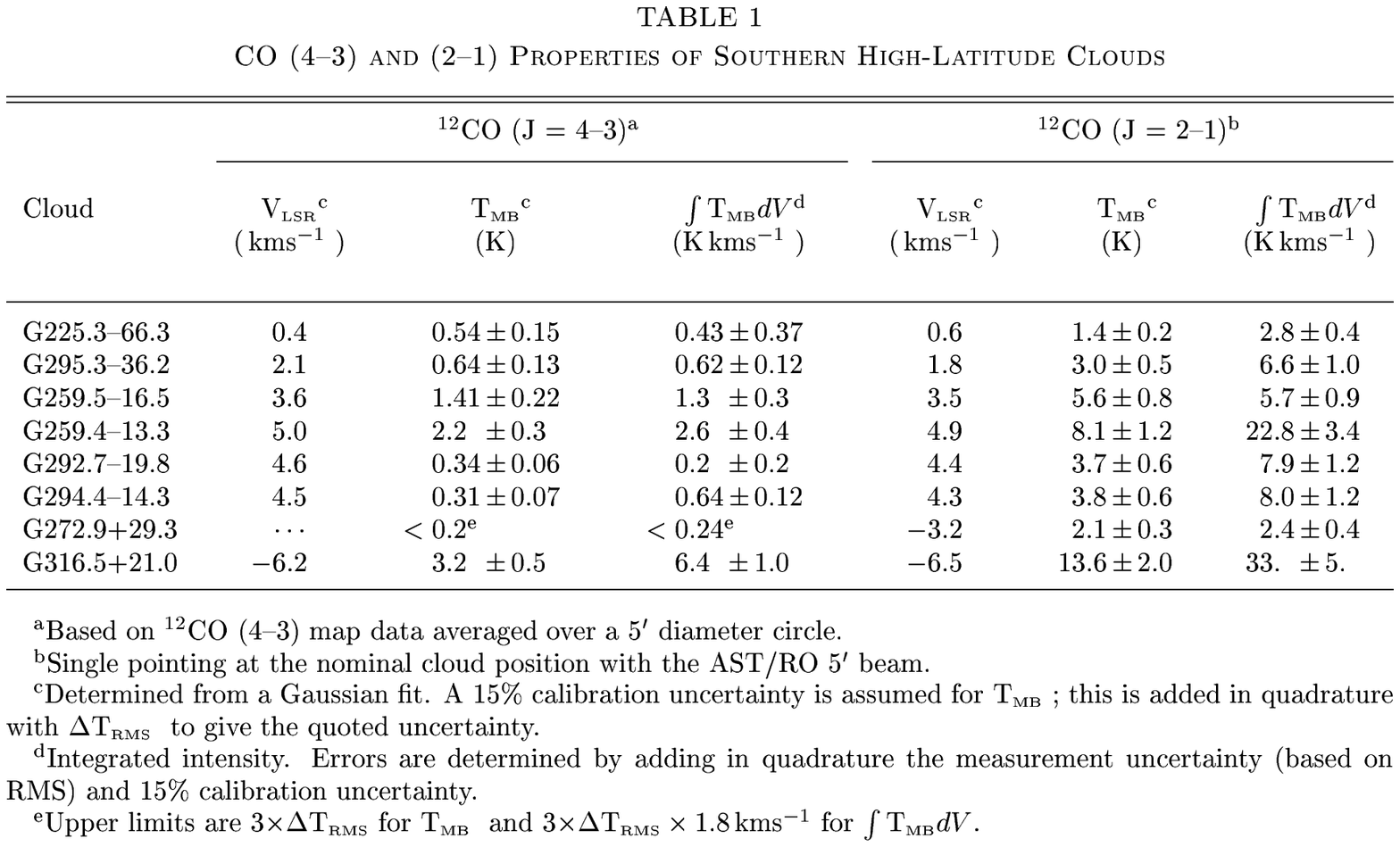}{4.0in}{0}{85}{85}{-255}{-190}
\end{figure*}

Given this geometrical uncertainty, data for many transitions are needed 
to decide if the primary source of CO emission is gas
with LDHT or HDLT conditions.  In particular, higher energy transitions of CO
should be observed towards translucent clouds.  Although they 
observed the (4--3) transition of \co\ towards some positions in the 
cloud they studied, FP were not able to determine 
if excitation of this transition was uniform, as it was for the (1--0) and
(2--1) transitions.  If the CO level populations are thermal, then
the discovery of uniform J = 4 excitation might support their claim of the 
ubiquity of gas with $n>\expo{4}\cmthree$, since
the critical density for thermal excitation of the (4--3) transition is 
$\sim\expo{4-5}\cmthree$.

\begin{figure*}[hb!]
\plotone{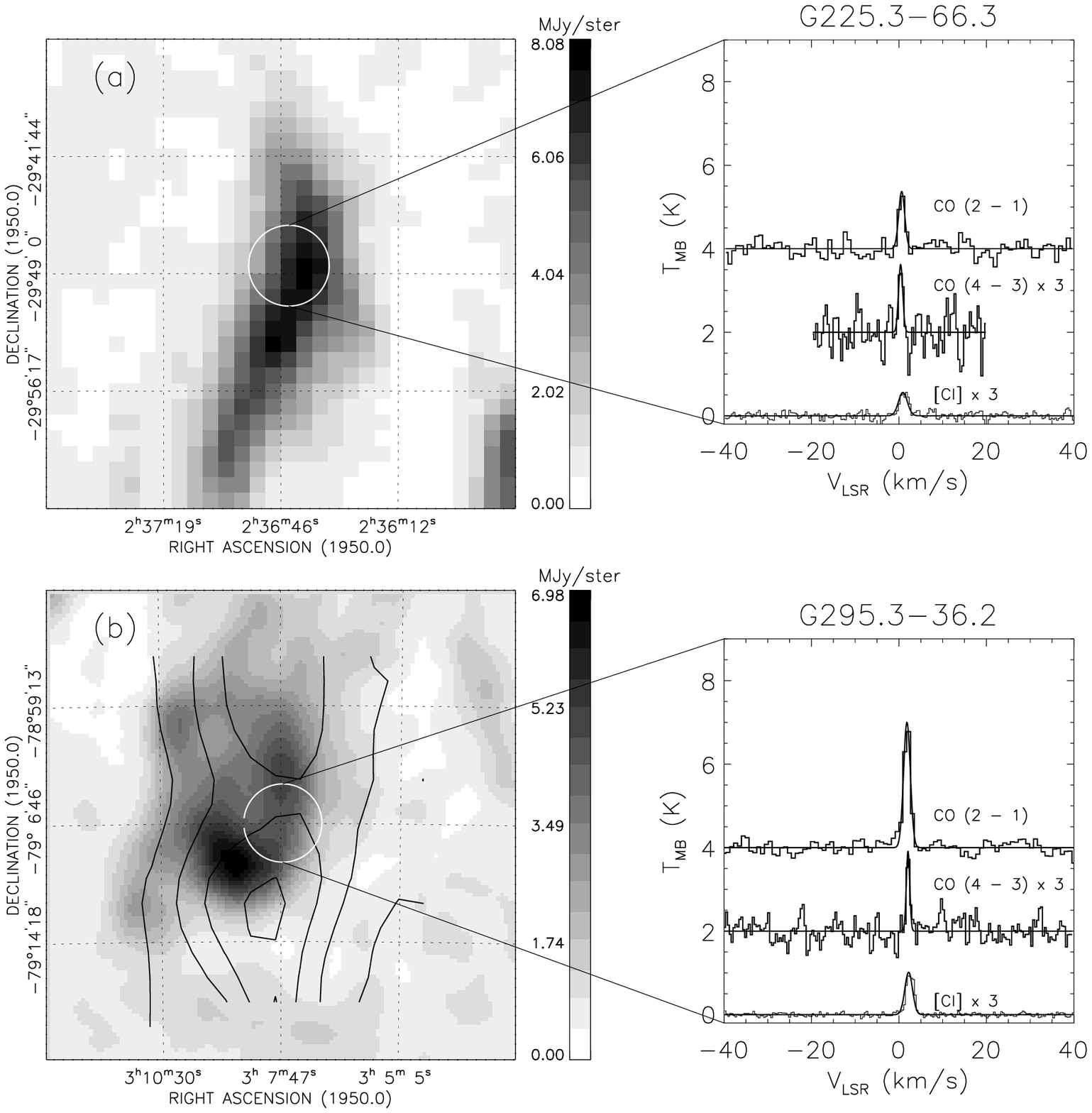}
\caption{Observations of high--latitude clouds.
The greyscale maps show HIRES-processed {\sl IRAS} 100\micron\
emission.  Black contours represent AST/RO \co\ (J = 2--1) 
integrated intensity.  A 5\arcmin\ circle is drawn on each map in white, 
at the nominal source position, defined by Keto \& Myers (1986).  The 
average \co\ (2--1), \co\ (4--3), and 
[\ci] (\pp) spectra measured in this circle with AST/RO are shown on the 
right.  Gaussian fits are superimposed on the spectra.  The \ci\ data were 
originally published by Ingalls \etal\ (1997).  The \co\ (4--3) and \ci\ 
intensities have been multiplied by 3.  ({\it a}) Cloud G225.3--66.3.  This
source was not mapped in \co\ (2--1) emission.
({\it b}) Cloud G295.3--36.2.  Contours of \co\ (2--1) emission range 
from 2.2 to 2.8 K\kms, in 2.2 K\kms\ intervals. \label{hlc1_2}}
\end{figure*}

In this paper we report observations of the J = (4--3) rotational 
transition of \co\ towards a sample of 8 translucent high--latitude 
molecular clouds at southern declinations (\S2 and \S3).  Our new data are 
used in conjunction with 
the \co\ and \cor\ observations made by van Dishoeck \etal\ (1991) to 
constrain the physical conditions in the HLC
regions emitting CO radiation (\S4).  We show that the emission from 
{\it all} CO transitions
up to (4--3) is linearly correlated, \ie\ the excitation
is uniform.  We use a statistical argument to demonstrate
that inclusion of our new (4--3) data narrows significantly the 
allowed region of parameter space, making the HDLT solution the most 
likely one.  
We discuss the physical implications of gas with HDLT properties, 
particularly thermal pressure imbalance with the ambient
interstellar medium (\S5).  Finally we suggest a method
to test for the existence of HDLT gas.

\section{Observations}

\subsection{The source sample}

Eight HLCs at southern declinations were observed (see Table 1 for Galactic
identifications).  Seven of the clouds were
listed in the catalog of Keto \& Myers (1986).  The observed source 
positions are listed in Ingalls \etal\ (1997).  In addition, a previously 
unknown translucent high--latitude cloud, G259.4--13.3 [$\alpha (1950.0)
= 7^h 29^m 42^s;~\delta (1950.0) = -47\arcdeg 00\arcmin 41\arcsec$], was observed.  This
cloud was discovered during a search made towards
{\sl IRAS} cirrus clouds for new \co\ (2--1) sources (Ingalls 1999).  

\begin{figure*}[ht!]
\plotone{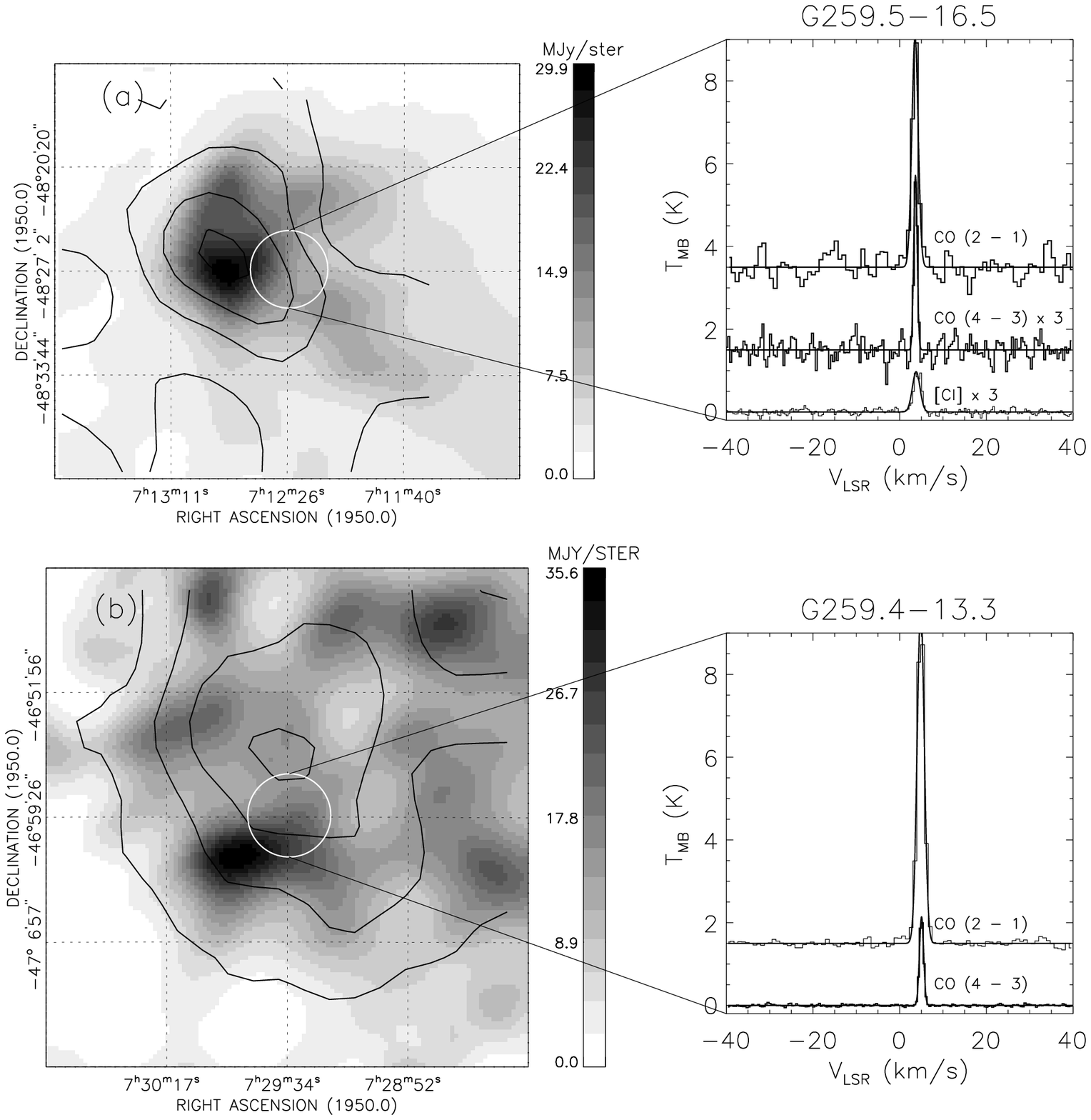}
\caption{Observations of high--latitude clouds.
  See caption to Figure \ref{hlc1_2} for description.  
  ({\it a}) Cloud G259.5--16.5.  Contours of \co\ (J = 2--1) emission range 
from 1.9 to 7.6 K\kms, in 1.9 K\kms\ intervals.  ({\it b}) Cloud G259.4--13.3,
discovered by Ingalls (1999), with nominal source position taken from that
work.  Contours of \co\ (2--1) emission range from 7.7 to 30.8 K\kms,
in 7.7 K\kms\ intervals. \label{hlc3_ib34}}
\end{figure*}
\begin{figure*}[hb!]
\plotone{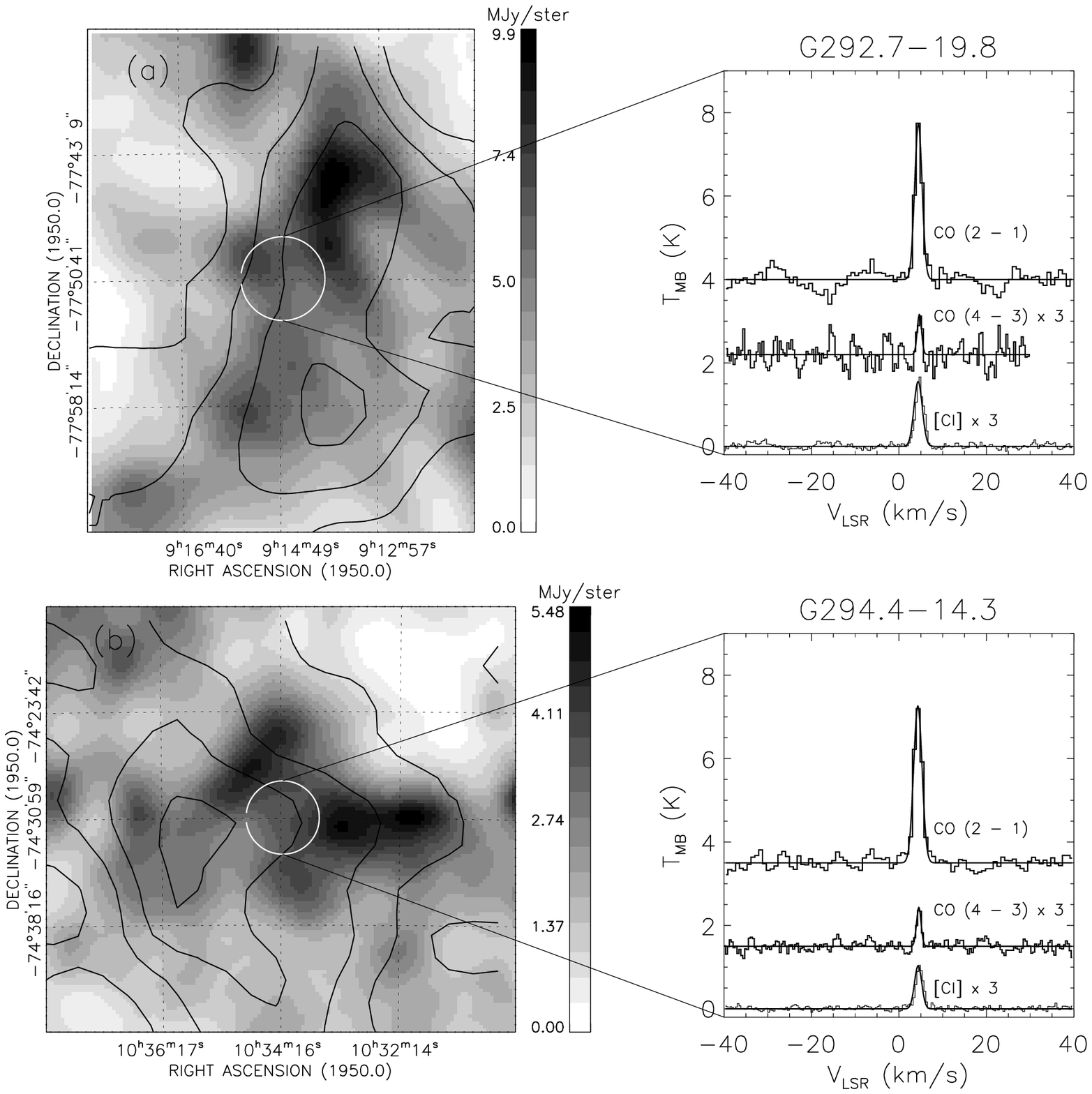}
\caption{Observations of high--latitude clouds.
  See caption to Figure \ref{hlc1_2} for description.  
  ({\it a}) Cloud G292.7--19.8.  Contours of \co\ (J = 2--1) emission range 
from 2.8 to 9.9 K\kms, in 2.4 K\kms\ intervals.  ({\it b}) Cloud G294.4--14.3.
Contours of \co\ (2--1) emission range from 2.4 to 9.9 K\kms,
in 2.5 K\kms\ intervals. \label{hlc4_6}}
\end{figure*}

The observations described here were part of a study of the overall gas-phase
carbon content in translucent clouds (Ingalls, Bania, \& Jackson 1994; 
Ingalls \etal\ 1997; Ingalls 1999).  The seven clouds 
from the Keto \& Myers (1986) catalog were each recently detected in
neutral atomic carbon [\ci ] emission by Ingalls \etal\ (1997).  Prior to this,
most of the clouds had only been observed in CO at 8\farcm 7 angular resolution
(Keto \& Myers 1986).  Cloud G225.3--66.3, has also been
observed at higher resolution in CO emission and in the photographic {\sl B} 
band by Stark (1995).  

Cloud G316.5+21.0 has a low mass star embedded in it 
(Ingalls \etal\ 1997).  Since this cloud has its own internal energy source
it may be hotter and more dense than the typical HLC.  A comparison with 
the more quiescent clouds in the sample may provide insight into
the effects of low mass star formation on translucent gas.

\subsection{AST/RO Observations of CO (4--3) and (2--1) emission}

\begin{figure*}[ht!]
\plotone{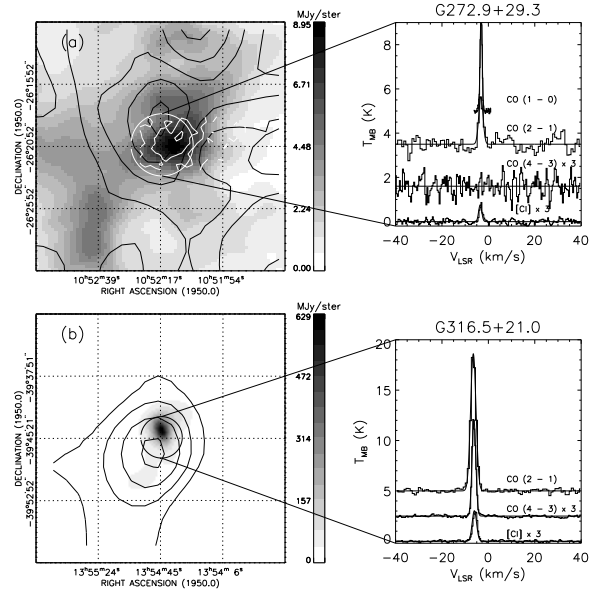}
\caption{Observations of high--latitude clouds.
  The data shown here correspond to those described in the caption
to Figure \ref{hlc1_2}, with the addition of FCRAO \co\ (1--0) data in part
{\it a}.    
  ({\it a}) Cloud G272.9+29.3.  White contours of \co\ (J = 1--0) emission 
range from 1.9 to 3.9 K\kms in 1.9 K\kms\ intervals.  The average (1--0) 
spectrum is superimposed on the plot at right.   Black contours of \co\ 
(J = 2--1) emission range 
from 0.6 to 3.0 K\kms in 0.6 K\kms\ intervals.  The \co\ (4--3) line was 
not detected from this source and the nondetection spectrum is shown.
  ({\it b}) Cloud G316.5+21.0.  Contours of \co\ (2--1) emission range from 
13.8 to 34.5 K\kms, in 6.9 K\kms\ intervals. \label{hlc7_8}}
\end{figure*}

Observations of the 461.041 GHz (J = 4--3) transition of \co\ were made
during the period 1997 June 29 through July 13, using the 
Antarctic Submillimeter Telescope and Remote
Observatory (AST/RO).  AST/RO is a 1.7m diameter telescope located at the 
National Science Foundation Amundsen-Scott South Pole Station (Stark
\etal\ 1997).  The telescope
efficiency at 461 GHz was estimated from skydip measurements (Chamberlin,
Lane, \& Stark 1997) to be $\eta_l = 0.77$.  As mentioned by Ingalls \etal\
(1997), $\eta_l$ is close to the main-beam efficiency, $\eta_{\litl MB}$.  
  The 461 GHz transition was measured using the AST/RO 492/810 SIS waveguide 
receiver which had a $\sim 130\K$ double sideband receiver 
noise temperature.  The intermediate frequency output was sampled by a
2048 channel, 1.4 GHz bandwidth acousto-optical spectrometer (AOS)
(Schieder, Tolls, \& Winnewisser 1989).  The velocity resolution
was $\delta$V$_{\litl 461} = 0.62\kms\,{\rm channel^{-1}}$.  System
noise temperatures were of order 1800--3000\K.  The half power beam width
of the AST/RO telescope at 461 GHz was approximately 180\arcsec.

Observations of the 230.538 GHz (J = 2--1) transition of \co\ were made in 
1996 November using the AST/RO 230 GHz SIS receiver which had a double
sideband noise temperature of $\sim 500\K$.  Based on measurements
of $\eta_l$ from skydips (see above), the value $\eta_{\litl MB} = 0.54$ 
was adopted for 230 GHz.  The same AOS used for the (4--3) measurements
was used for the (2--1) data, yielding a velocity resolution of 
$\delta$V$_{\litl 230} = 1.34\kms\, {\rm channel^{-1}}$.  The system noise
temperature at 230 GHz was $\sim 1800\K$.  The beamsize of the AST/RO
telescope at 230 GHz is $\sim 5\arcmin$.

The \co\ (4--3) and (2--1) observations were obtained by 
position-switching 1\arcdeg\ in azimuth.  Because of the location of AST/RO,
position-switched offsets were both at constant elevation 
and at fixed positions on the sky.  Intensity calibration was accomplished 
by measuring
two blackbody loads at known temperatures (Stark \etal\ 1997).  In order
to correct for atmospheric attenuation the sky brightness temperature 
was measured at the position of a source every $\sim 15$ minutes.  
The overall intensity calibration was checked by observing 
periodically the compact \hii\ region G291.2--0.8.  This procedure yielded 
intensities repeatable to within $\pm 15\%$ at both 230 GHz and 461 GHz.  

\subsection{FCRAO observations of CO (1--0) towards cloud 
G272.9+29.3}
The HLC G272.9+29.3 was observed with the Five
College Radio Astronomy Observatory (FCRAO), in Amherst MA, on 1996 
February 17 through 18.  The FCRAO 14m telescope was used to map the \co\ 
(1--0) transition at 115.2712 GHz.  The telescope has a beamsize of
45\arcsec\ and a 115 GHz main beam efficiency of $\eta_{\litl MB} \approx 0.36$
at 20\arcdeg\ elevation (Ladd \& Heyer 1996).  The observations 
were carried out in
frequency--switched mode, with the signal and reference spectra spaced
4.5 MHz apart.  The spectra were measured with the QUARRY focal plane array
receiver (Erickson \etal\ 1992), and the data were sampled by an array of 15 
autocorrelation spectrometers with 1024 channels, separated by 77 kHz per 
channel, giving a velocity resolution of $\delta$V$_{\litl 115} = 0.2\kms\,
{\rm channel^{-1}}$.  Since the source is
a Southern hemisphere object with declination $\delta = -26\arcdeg$, its
elevation was rather low, ranging from 16\arcdeg\ to 19\arcdeg.  
The single--sideband system noise temperature was approximately
2000 K for these observations.

\section{Observational Results}

\subsection{CO (4--3) and (2--1) Properties of HLCs}

Observed \co\ (4--3) and (2--1) spectra are 
plotted alongside maps of each source in Figures \ref{hlc1_2} to 
\ref{hlc7_8}.  The (4--3) transition
was detected towards 7 of 8 clouds observed.  
The \co\ (4--3) HLC data consist of half-beam sampled 
measurements made on a $\sim 5\arcmin \times 5\arcmin$ grid,
in a manner similar to the \ci\ ``superbeam'' observations described by
Ingalls \etal\ (1997).  These maps were averaged over the 
$5\arcmin$--square grids to
produce the spectra shown in Figures \ref{hlc1_2} to \ref{hlc7_8}.  The 
\ci\ ``superbeam'' average spectra taken towards the same positions 
(Ingalls \etal\ 1997) are also displayed in the figures.  Intensities are 
main-beam
brightness temperatures, \tmb, \ie\ Rayleigh-Jeans antenna temperature,
$T_A$, corrected for both atmospheric attenuation and main-beam efficiency
($\tmb \equiv T_A^*/\eta_{\litl MB}$).

Each source was observed in \co\ (2--1) emission at its nominal 
center with the $\sim 5\arcmin$ AST/RO beam.  For seven 
of the sources, these spectra are part of\ $\sim\,30\arcmin\times
30\arcmin$ half-beam sampled maps, shown as black 
contours overlaid on {\sl IRAS} 100$\,\mu$m greyscale images in Figures 
\ref{hlc1_2} to \ref{hlc7_8}.  The {\sl IRAS} images were all processed using 
the HIRES technique (Levine \etal\ 1993), and have angular resolution 
$\sim 100\arcsec$.  The calibration uncertainty for the HIRES images is 
of order 20\%.  

Gaussian fits were made to the CO (4--3) and (2--1) spectra and are
shown superimposed on these spectra in Figures \ref{hlc1_2} to 
\ref{hlc7_8}.  Table 1 lists the line center
velocity, \vlsr, and the peak intensity, \tmb, from the Gaussian fits.  Also
listed is the integrated line intensity, $\int\tmb dV$.  Error bars for 
both transitions were 
determined by adding in quadrature the instrumental noise (\dtrms) and a
calibration uncertainty of 15\%:
\begin{equation}
\sigma(\tmb) =  \sqrt{\left(\dtrms\right)^2 + \left(0.15\tmb\right)^2}.
\label{error_bar}
\end{equation}
Gaussian fit linewidths, $\Delta V$, range from 0.9 to 3\kms\ with an average
of $(1.7 \pm 0.5)$\kms, slightly greater than the velocity resolution of 
the (2--1) observations, so the lines are only marginally resolved.  For 
this reason both \tmb\ and $\Delta V$ 
are poorly-determined.  We will therefore focus our attention on the integrated
intensity, $\int\tmb dV$, which is a better defined quantity.  Since 
(4--3) radiation was not detected towards
G272.9+29.3, we estimate a 3$\sigma$ upper limit to the (4--3) integrated 
intensity for this source using the average linewidth:  $\int\tmb(4-3)\,dV <
3\times\dtrms\times 1.7\kms$.

\subsection{The CO (1--0) Map of G272.9+29.3}
The (1--0) transition of \co\ was mapped in source G272.9+29.3 at 
45\arcsec\ angular resolution.  The map consisted of a $29 \times 11$
grid of pixels, evenly spaced at $\sim 25\arcsec$, one-half the FCRAO
beamwidth.  The resulting map of \co\ (1--0) integrated intensity is 
superimposed as
white contours on Figure \ref{hlc7_8} ({\it a}).  The average spectrum in a 
$5\arcmin$ ``superbeam'' at the nominal source center is plotted on the right 
side of this figure.  This spectrum was resolved in
velocity by the FCRAO spectrometer:  a
Gaussian fit gives the line center velocity $\vlsr = -3.1$, linewidth
$\Delta V = 1.0$ and peak intensity $\tmb = 5.6\pm 1.8$.  

\begin{figure*}[ht!]
\plotonesmall{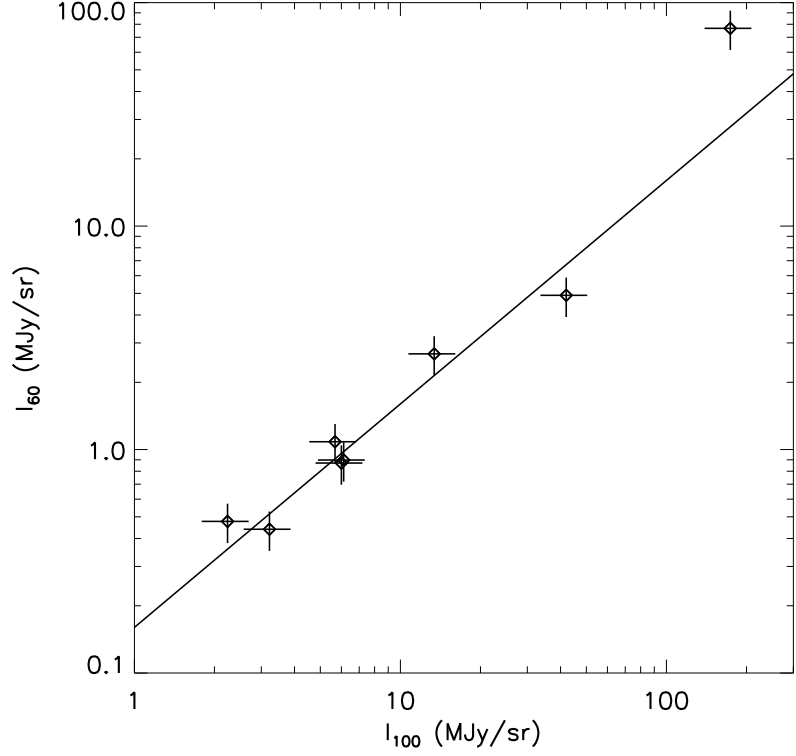}
\caption{The HLCs compared with typical Galactic emitters of infrared 
radiation.
  This plot shows the 60\micron\ versus 100\micron\ surface brightness of 
the sample clouds.  A straight line denoting $\irasc = 0.16\irasd$, the average
behavior for Milky Way gas correlated with \hi\ emission (Dwek \etal\ 1997), 
is superimposed.  Note that this line is {\it not} a fit to the data.
\label{iras}}
\end{figure*}

\subsection{The HLCs as Representative CO Clouds}
The HLCs are the average Galactic high--latitude sources of CO emission.  
  This can be shown in two ways.  First, we can use the all-sky CO
survey done with the Far Infrared Absolute Spectrometer (FIRAS) instrument
on board the {\it Cosmic Background Explorer} ({\sl COBE}) satellite.  This
instrument was used to observe
nearly the entire sky at 7\arcdeg\ resolution in the first six 
transitions of \co\ (Wright \etal\ 1991;
Bennett \etal\ 1994).  The data are available on the World Wide 
Web.\footnote{The Web site is http://www.gsfc.nasa.gov/astro/cobe/.  The 
{\sl COBE} datasets were developed by 
NASA's Goddard Space Flight Center under the guidance of the {\sl COBE} 
Science Working Group and were provided by the National Space Science Data
Center.}  The average value of the {\sl COBE} FIRAS high-latitude 
($|b| > 15\arcdeg$) (4--3)/(2--1) integrated intensity ratio is
$\langle (4-3)/(2-1)\rangle_{\litl COBE} = 0.84\pm 0.06$.  The intensities 
were reported in units of erg\cmtwo\persec\,${\rm sr}^{-1}$.  The 
FIRAS integrated intensity ratio may be compared with the ratio of integrated 
Rayleigh-Jeans brightness temperatures measured here for HLCs if one uses 
the conversion formula:
\begin{equation}
\frac{\int I_{\nu} d\nu}{\rm erg\cmtwo\persec sr^{-1}}~=~
Q(\nu)\,\frac{\int \tmb\,dV}{\rm K\kms}.
\label{conversion}
\end{equation}
The function $Q(\nu) \equiv 2\times 10^{5} k \nu^3/c^3$:  $Q(230) =
1.25\times 10^{-8}$ and $Q(461) = 10^{-7}$\,erg\cmtwo
\persec\,${\rm sr}^{-1}/(\K\kms)$.  The average HLC value measured here is 
[Equation \ref{co_fit}, multiplied by $Q(461)/Q(230) = 8$]
$\langle (4-3)/(2-1)\rangle_{\litl HLC} = 0.88\pm 0.16$, which
is nearly identical to the {\sl COBE} average. 

\begin{figure*}[ht!]
\plotonesmall{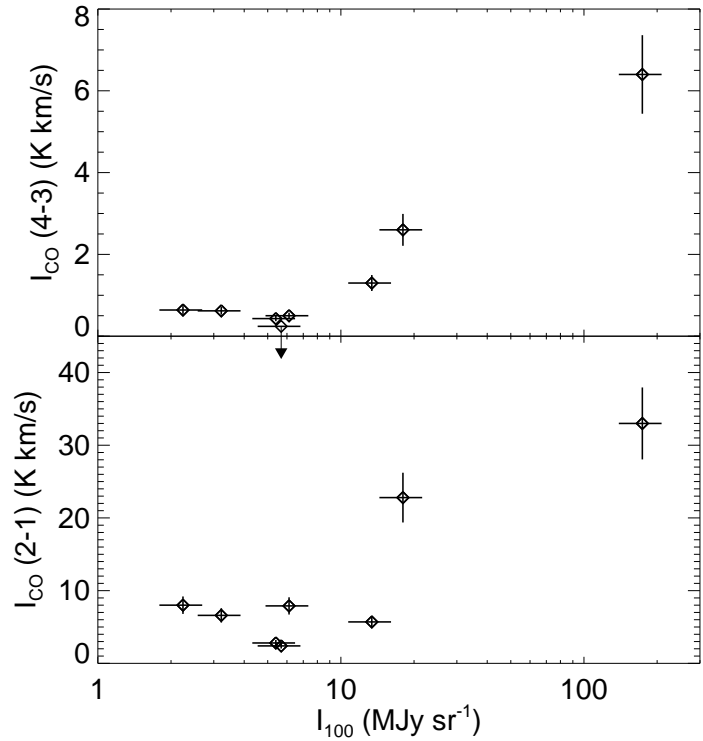}
\caption{Observed \co\ (J = 4--3) ({\it Top}) and (J = 2--1) ({\it Bottom})
properties of HLCs.  The
integrated main beam brightness temperature, $\int\tmb dV$, is plotted here
for each transition as a function of {\sl IRAS} 100\micron\ surface 
brightness. \label{colines_1}}
\end{figure*}
\begin{figure*}[hb!]
\plotonesmall{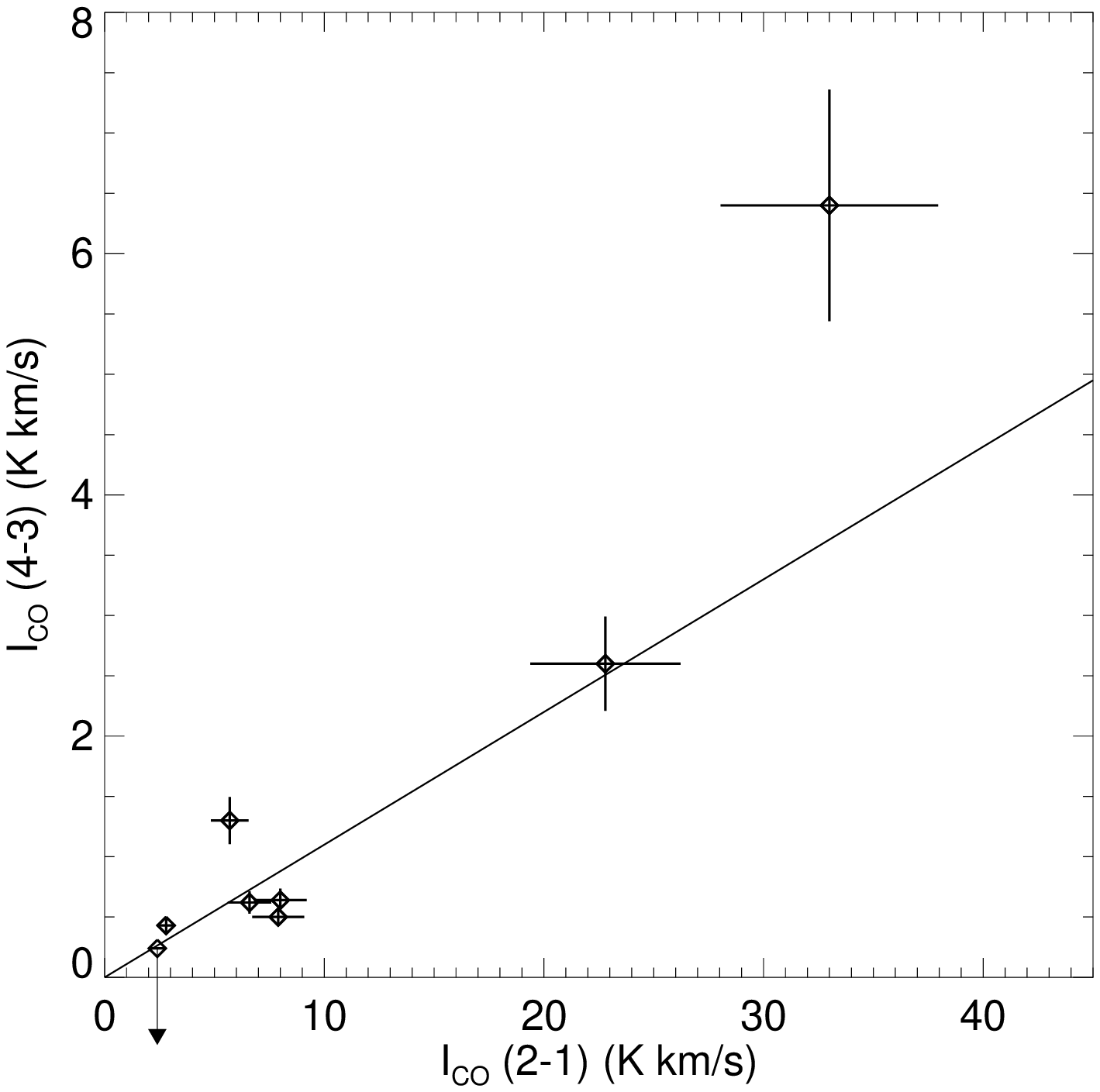}
\caption{Integrated main beam brightness temperature \ico\ (4--3) versus
\ico\ (2--1).  The straight line fit, $I_{(4-3)} = 0.11I_{(2-1)}$, 
is superimposed. \label{colines_2}}
\end{figure*}

The infrared properties of the HLCs studied here can also be 
compared with the average properties of interstellar cirrus clouds.  
HIRES-processed images of both 60 and 100\micron\ emission observed with 
{\sl IRAS} were obtained for each cloud in the sample.  The 
60\micron\ surface brightness, \irasc,
is plotted as a function of the 100\micron\ surface brightness, \irasd, in 
Figure \ref{iras}.  Both the 60
and 100\micron\ measurements are averages over the 5\arcmin\
``superbeam'' centered on each source, described in \S3.1.  The Diffuse 
Infrared Background Experiment (DIRBE) on board the {\sl COBE}
 satellite was used to observe the infrared spectrum of the entire Galaxy.  
The average result for all Milky Way gas correlated with \hi\ emission,
$\langle \irasc /\irasd\rangle_{\litl COBE} = 0.160$, was derived by Dwek 
\etal\ (1997).  
This is the same as the mean value for the HLCs studied here,
$\langle \irasc /\irasd\rangle_{\litl HLC} = 0.156 \pm 0.016$ (the authors
did not quote an error for this ratio).  A straight 
line indicating $\irasc = 0.16\irasd$ is superimposed on Figure \ref{iras}. 
Notice that most of the clouds fall on this line, with the exception of source 
G316.5+21.0 which has $\irasc = 77\pm 15\mjysr$.  For this cloud $\irasc /\irasd = 0.44 \pm 0.12$, which is much 
larger than that of the other clouds in the sample.  This implies that 
G316.5+21.0 has a greater relative 
abundance of the ``very small grains'' which produce most of the 60\micron\ 
emission (\eg\ D\'esert, Boulanger, \& Puget 1990).  The 
low mass star which is embedded in the source must be responsible for this,
presumably by heating the dust in excess of the interstellar radiation 
field (ISRF) and evaporating the mantles of larger grains which 
produce 100\,\micron\ emission.  

\subsection{Correlations Between Integrated Intensities}

Here we give evidence for uniform CO excitation in translucent clouds.  
Figure \ref{colines_1} shows the observed (4--3) and (2--1) integrated
intensities for the HLC sample, plotted as a function of \irasd .  
Both graphs in Figure \ref{colines_1} show the same behavior, although 
the trend is not necessarily monotonic.  Figure 
\ref{colines_2} is a plot of $I_{(4-3)}
\equiv\int\tmb(4-3)\,dV$ as a function of $I_{(2-1)}\equiv
\int\tmb(2-1)\,dV$.  The (4--3) integrated intensity is linearly correlated 
with the (2--1) integrated intensity.  The correspondence is
significant, with a correlation coefficient of $r = 0.95$.  Linear 
regression to the data, weighted by the errors, gives:
\begin{equation}
I_{(4-3)}~~ = ~~ (0.11\pm 0.02)\,I_{(2-1)}.
\label{co_fit}
\end{equation}
A line representing Equation \ref{co_fit} is
superimposed on the data in Figure \ref{colines_2}.  The 
$I_{(4-3)}$--intercept was held fixed at zero, \ie\ we have
assumed that there is no (4--3) emission when there is no (2--1)
emission, and vice--versa.  Source G316.5+21.0 has an anomalously 
high (4--3)/(2--1) integrated intensity ratio.  As explained in \S3.3,
the star embedded in this cloud is likely to make the gas hotter
than in the average translucent cloud, whence high level CO transitions 
will be excited more readily into emission.  Since the linear fit to the 
(4--3) and (2--1) data was weighted by the errors, this source (which had
rather large error bars) did not contribute significantly to the fit.

The linear relationship between $I_{(4-3)}$ and $I_{(2-1)}$ 
(Figure \ref{colines_2}) 
is reminiscent of the correlation between the 
\co\ (2--1) and (1--0) brightness temperatures seen towards
``pre-star-forming regions'' by Falgarone \etal\ (1998).  There is nothing
special about the HLC positions we observed; they are the nominal 
``cloud center'' positions (Keto \& Myers 1986; Ingalls 1999), and do not
necessarily coincide with the peak of CO emission.  Recall
that all quantities quoted in Table 1 refer to measurements in a 5\arcmin\ 
circle (see Figures \ref{hlc1_2} to \ref{hlc7_8}) made 
towards these centers.  These nominal positions have 100\micron\ surface 
brightnesses ranging
over 2 orders of magnitude, so the sample is obviously not homogeneous.  
Nevertheless, the ratio between the (4--3) and (2--1)
integrated intensities, as well as between the 60 and 
100\micron\ intensities, is remarkably uniform.  The simplest 
interpretation is that {\it the CO emission from all translucent clouds 
originates
from gas with the same physical conditions}, even though different lines 
of sight contain different amounts of such gas.  This was the conclusion
of Falgarone \etal\ (1991) and FP, based on their
study of the edge of a cloud in the Perseus complex.  

\begin{figure*}[ht!]
\plotonesmall{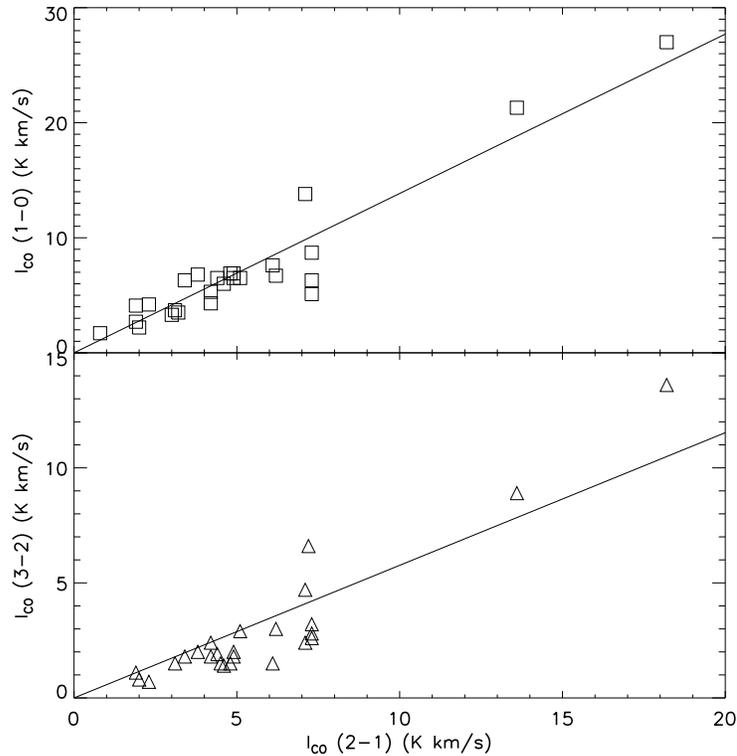}
\caption{Integrated intensity of the \co\ (1--0) ({\it Top}) and (3--2) 
({\it Bottom}) transitions, plotted as a function of the (2--1) integrated
intensity for translucent and high--latitude clouds (from
Table 7 of van Dishoeck \etal\ 1991).  The linear fits to these data, 
$I_{(1-0)} = 1.3I_{(2-1)}$ and $I_{(3-2)} = 0.55I_{(2-1)}$, are
superimposed.\label{cofits}}
\end{figure*}

A linear correlation seems to exist between {\it all} pairs of CO transitions 
observed towards HLCs.  The most extensive database of CO observations 
of HLCs is that of van Dishoeck \etal\ (1991).  Figure \ref{cofits} shows
$I_{(1-0)}$ ({\it Top}) and $I_{(3-2)}$ ({\it Bottom})
plotted as a function of $I_{(2-1)}$ for the $\sim 25$ translucent and/or 
high--latitude positions detected by van Dishoeck \etal\ (1991; their 
Table 7).  Both 
the $I_{(1-0)}$ vs. $I_{(2-1)}$ and the $I_{(3-2)}$ vs. $I_{(2-1)}$ datasets
have a highly significant correlation ($r = 0.94$ for each dataset),
of which van Dishoeck \etal\ (1991) were apparently unaware.  Linear 
regression to the data gives the following fits:
\begin{equation}
I_{(1-0)} ~~ = ~~ (1.3 \pm 0.4)\,I_{(2-1)};
\label{co_fit_b}
\end{equation}
\begin{equation}
I_{(3-2)} ~~ = ~~ (0.55\pm 0.16)\,I_{(2-1)}.
\label{co_fit_c}
\end{equation}
As for Equation \ref{co_fit}, the fits were constrained to have an
intercept value of zero.  We weighted the data using 30\% error bars, 
corresponding to the 
overall calibration uncertainty quoted by van Dishoeck \etal\ (1991).  This 
correlation also applies to the cloud sample used by Turner (1993).  For 
Turner's translucent cloud sample, the average \co\ (1--0)/(2--1) integrated 
intensity ratio for nine sources is $1.6\pm 0.3$, which is indistinguishable 
(within the errors) from the result given above.

\begin{figure*}[hb!]
\plotonesmall{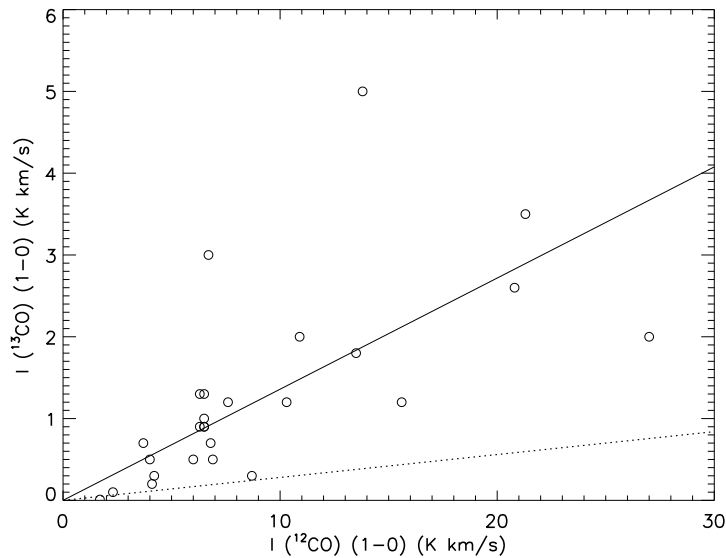}
\caption{Integrated \cor\ (1--0) intensity versus integrated \co\ (1--0)
intensity for translucent and high--latitude clouds (Table 7
of van Dishoeck \etal\ 1991).  The linear fit 
$I({\rm ^{13}CO}) = 0.13I({\rm ^{12}CO})$ is superimposed as a solid
line.  The upper limit behavior expected for optically thin emitting regions, 
$I({\rm ^{13}CO}) = 0.028I({\rm ^{12}CO})$ (see text), is superimposed as 
a dashed line.  In general the \co\ (1--0) emission is optically
thick.\label{cofits_b}}
\end{figure*}

The slope in Equation \ref{co_fit_b} is equivalent to the constant 
(2--1)/(1--0) brightness temperature ratio observed by
Falgarone \etal\ (1998).  Considering each spectral
channel individually, they measured $T_{(2-1)}/T_{(1-0)} = 
0.65\pm 0.15$ for 80\% of the data points in their maps made towards 
three translucent clouds.  In comparison,
the reciprocal of the Equation \ref{co_fit_b} slope is 
$I_{(2-1)}/I_{(1-0)} = 0.77\pm 0.24$.  Note that Falgarone \etal\ (1998) and 
FP did not study isolated translucent objects, but since they obtained similar 
values for $I_{(2-1)}/I_{(1-0)}$ it is possible that the linear relationship 
is also a characteristic of the {\it translucent edges} of non-starforming dark clouds.

This result is not limited to translucent regions.   A large 
scale survey of the Milky Way molecular ring has obtained an average
value of the (2--1)/(1--0) ratio of $0.85\pm0.63$ with a median of 0.69 
(Chiar, Kutner, Verter, \& Leous 1994).  Similarly, the mean 
value for the entire Galactic plane,
$\langle T_{(2-1)}/T_{(1-0)}\rangle = 0.77$, was derived by 
Sanders \etal\ (1993; also see Sanders \etal\ 1986), and Handa \etal\ (1993) 
measured a mean ratio of 0.7 for the Galactic molecular ring (also see
Dame \etal\ 1986).  Thus essentially all surveys of CO consistently
show a linear correlation between (2--1) and (1--0) integrated emission.

This correlation also exists for the CO peak brightness temperature.  
Peak brightness temperatures were listed by van Dishoeck \etal\ (1991)
for the translucent sources with spectrally resolved (1--0), (2--1), and 
(3--2) line profiles.  We find that these are also 
correlated, and linear fits give the same slopes as the integrated 
intensity fits.  In general, for any two of these transitions, ``1'' and ``2'',
\begin{equation}
\frac{\int\tmb(1)\,dV}{\int\tmb(2)\,dV} ~ = ~ \frac{\tmb(1)}{\tmb(2)}
~~~,
\label{tmb_tint}
\end{equation}
This is a trivial statement for Gaussian line profiles but according
to van Dishoeck \etal\ (1991) not all of their spectra were well-described
by a Gaussian shape.  Therefore,
linear correlation of integrated intensities is not solely due 
to a correlation in {\it linewidths} which might 
occur in optically thick gravitationally bound systems with different 
masses.  

We make one last comparison, that between the integrated intensities of
the (1--0) transition of the \cor\ and \co\ isotopomers.  The data are 
listed in Table 7 of van Dishoeck \etal\ (1991) and plotted here in 
Figure \ref{cofits_b}.  For 
these data the correlation coefficient, $r=0.64$, so there is only an 
80\% chance that the data are 
correlated (Taylor 1997).  Linear regression (with zero intercept) yields:
\begin{equation}
I({\rm ^{13}CO}) ~~ = ~~ (0.13\pm 0.04)I({\rm ^{12}CO}).
\label{co_fit_13}
\end{equation}
The fit is not very robust, but the slope 0.13$\pm$0.03 gives a
good estimate of the average \cor/\co\ intensity ratio.  (We obtain
a comparable value for this ratio using the \cor\ and \co\ HLC data of Keto
\& Myers 1986).  This is approximately five times higher than what is
expected if the
CO-emitting regions are optically thin to both \cor\ and \co\ radiation,
assuming the abundance ratio $[^{12}{\rm C}]/[^{13}{\rm C}]= 60$ 
(Boreiko \& Betz 1996; Langer \& Penzias 1993).  Taking into account 
isotope-selective
photodissociation and chemical fractionation, the slope from optically thin
gas should range from $I(^{12}{\rm CO})/I(^{13}{\rm CO})\approx
 0.014$ to 0.028 (van Dishoeck \& Black 1988).  
Nearly all of the data points fall above the line
$I(^{13}{\rm CO}) = 0.028I(^{12}{\rm CO})$ which is superimposed on Figure
\ref{cofits_b}, so we conclude that most of the CO-emitting regions in
HLCs are optically thick to \co\ (1--0) radiation.  

We have shown that the CO integrated intensities $I_{(1-0)}$, $I_{(2-1)}$, 
$I_{(3-2)}$, and $I_{(4-3)}$ are related linearly for a large sample of 
translucent high--latitude molecular clouds.  Unfortunately the cloud
sample used by van Dishoeck \etal\ (1991) does not overlap with our own 
sample, so it is uncertain whether all of the clouds in
both samples fulfil all of the linear relationships.  This is probably true,
however, since both samples probe the same kind of gas.  By definition, 
all of the clouds are translucent (\ie\ $\av = 1-5$, no internal UV 
sources; see Ingalls \etal\ 1997).  In addition, they have similar \co\ (2--1) and 
100\micron\ emission 
properties.  For example, the weighted mean $I_{(2-1)}/\irasd$ ratio 
for our sample is $0.55\pm 
0.06\K\kms\,(\mjysr)^{-1}$.  For the cloud towards HD 210121,
towards which van Dishoeck \etal\ (1991) observed six positions,
the average $I_{(2-1)}/\irasd$ ratio is $0.6\pm 0.4\K\kms\,(\mjysr)^{-1}$.  It
is therefore likely that the clouds in both samples are equivalent.  Indeed, 
Falgarone \etal\ (1998) cite the linear 
relationship between intensities of CO rotational transitions as a 
general characteristic of ``clouds of low average column density'', 
\ie\ translucent clouds (see also Clemens \& Barvainis 1988;
Falgarone \etal\ 1991; Falgarone, Puget, \& P\'erault 1992; FP).  

In what follows we interpret the slopes of the linear fits 
(Equations \ref{co_fit}, \ref{co_fit_b}, \ref{co_fit_c}, and 
\ref{co_fit_13}) to be the true (constant) values of the respective 
integrated intensity ratios in the cells which emit CO radiation.  We take 
the CO-emitting cells to be nearly identical structures 
which are separated from each other in position and/or radial velocity.  In 
this model an increase in the integrated intensity of a CO transition 
is attributed solely to an increase 
in the number of cells being observed, \ie\ an increase in $\phi$, the 
beam filling fraction of the CO radiation emitting from a particular
direction at a specific velocity.  

\begin{figure*}[ht!]
\plotonesmall{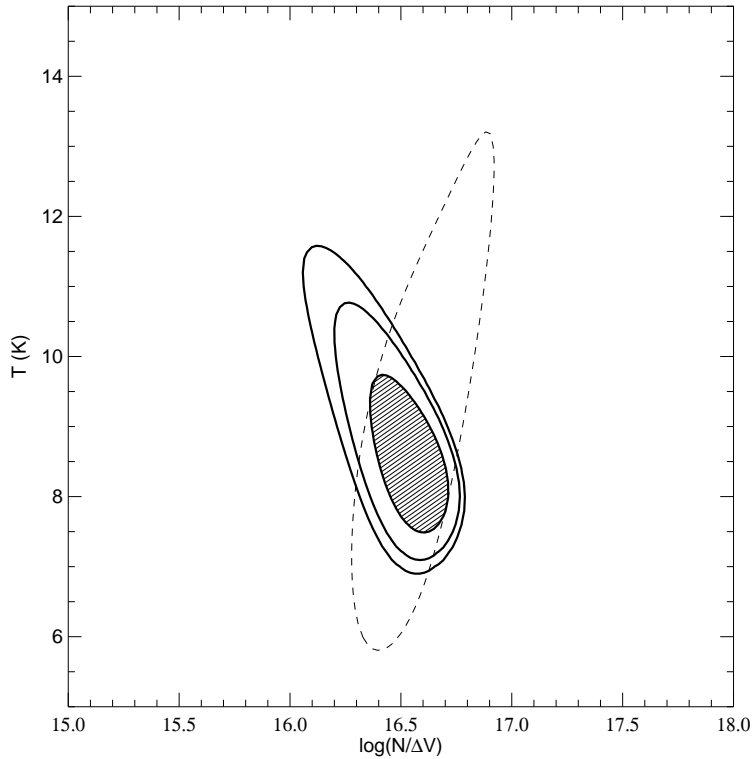}
\caption{Probability analysis of HLCs in ($\nco/\Delta V$)--temperature space, 
under LTE conditions (see Table 2).  The 
dashed curve encloses the 1$\sigma$ confidence region when only HLC data 
for the (3--2) and lower transitions of \co\ and
\cor\ are taken into account.  The solid contours enclose the 
1,2, and 3$\sigma$ confidence intervals when the new (4--3)/(2--1) data 
(this paper) are included.  The inner, 1$\sigma$, contour
is shaded.  The units of $\nco/\Delta V$
are \cmtwo\,$(\kms)^{-1}$.  \label{cloud_lte}}
\end{figure*}

\null\vskip.5in
\section{Modelling the CO-emitting regions:  A statistical approach}
Here we introduce a statistical method for modelling the physical conditions
in the CO-emitting cells of translucent molecular clouds.  As stated
above, integrated intensity ratios among the first four 
transitions of CO are constant.  Define the following
integrated intensity ratios for translucent clouds:
\begin{eqnarray}
\frac{I_{(2-1)}}{I_{(1-0)}} \equiv  (R_1\pm \sigma_1) & = & 0.77\pm 0.24;
\label{R1}\\ \nonumber\\
\frac{I_{(3-2)}}{I_{(2-1)}} \equiv (R_2\pm \sigma_2)&  = & 0.55\pm 0.16;
\label{R2}\\ \nonumber\\
\frac{I_{(4-3)}}{I_{(2-1)}} \equiv  (R_3\pm \sigma_3)& = & 0.11\pm 0.02;
\label{R3}\\ \nonumber\\
\frac{I[^{13}{\rm CO}(1-0)]}{I[^{12}{\rm CO}(1-0)]}\equiv (R_4\pm \sigma_4) & = &
0.13\pm 0.04\label{R4}.  
\end{eqnarray}
These ratios are taken from Equations \ref{co_fit}, \ref{co_fit_b},
\ref{co_fit_c}, and \ref{co_fit_13}.  Suppose that fluctuations in the 
$i$th ratio, $r_i$, are distributed normally with mean $R_i$
and standard deviation $\sigma_i$.  The probability of
measuring the $i$th ratio to be between $r_i$ and $r_i+dr_i$ is then:
\begin{equation}
p(r_i,R_i,\sigma_i)dr_i ~ = ~ \frac{1}{\sigma_i\sqrt{2\pi}}{\rm e}^{-(r_i-R_i)^2/(2\sigma_i^2)}\,dr_i.
\label{gaussian}
\end{equation}
Since the ratios are constant (excluding fluctuations), they are
statistically independent of each other.  The joint
probability of observing a given set of ratios is thus given by the product:
\begin{equation}
P(r_1,r_2,r_3,r_4)dr_1dr_2dr_3dr_4\equiv P({\bf r})d{\bf r} = \prod_{i=1}^4 p(r_i,R_i,\sigma_i)dr_i.  
\label{prob_dist}
\end{equation}

This analysis requires that all the CO emission originates in gas with the 
same physical conditions.  We of course have argued that this is the
case (see \S3.4).  Therefore, given a physical model and an input grid of 
physical parameters, the line intensities, \tmb, and ratios among 
them, $r$, can be computed.  Equation \ref{prob_dist} gives the 
probability distribution in the vector of ratios,
{\bf r}.  This can be used to estimate the most likely portion of 
the input parameter space which gives the observed ratios.  Below we 
evaluate the high density, low temperature (HDLT) and the low density, high
temperature (LDHT) translucent cloud scenarios using two simple 
cloud models, the local thermodynamic equilibrium model and the 
large velocity gradient model.
\subsubsection{The LTE model}\nobreak
In the HDLT scenario advocated by FP the CO emission 
originates in cold, high-density regions where the CO level populations are 
in local thermodynamic equilibrium (LTE).  This scenario is worth 
investigating because of its extreme simplicity---in LTE the line ratios 
depend on only two parameters, the excitation temperature, \tex , and the 
CO column density per velocity interval, \ncodv.


The peak intensity of a CO spectral line, \tmb, is given in LTE by:
\begin{eqnarray}
{\tmb} & = & 5.53\K\,\left[ \frac{1}{{\rm e}^{(5.53\K)J/\tex} -1} -
\frac{1}{{\rm e}^{(5.53\K)J/{\rm T_{BG}}} -1} \right] \nonumber
\\ \nonumber \\ 
& & (1 - {\rm e}^{-\tau})\,\phi,\label{tmb_lte} 
\end{eqnarray}
where ${\rm T_{\litl BG}} = 2.7\K$ is the cosmic microwave background 
temperature and $\tau$ is the line center opacity.  The transition is 
from rotational level $J$ to $J-1$.  The function $\phi$ contains the
details of the telescope beam average; if the emission originates 
from gas with uniform properties, then $\phi$ is simply the beam filling 
fraction of the gas.  In this case, which we assume to hold for HLCs, $\phi$ 
cancels when line ratios are taken.  

We will use the 
LTE model to predict \tmb\ ratios, and then compare them directly with the 
observed $\int\tmb\,dV$ ratios (Equation \ref{tmb_tint}) by 
assuming Gaussian line profiles.  The LTE opacity for a transition from 
level $J$ to level $J-1$ is given by: 
\begin{eqnarray}
\tau_{J,J-1}  &=&  \frac{8\pi^3\mu^2}{3hZ(\tex)}\,{\rm e}^{-(2.765\K)J(J+1)/\tex} \nonumber \\
& & [{\rm e}^{(5.53\K)J/\tex} - 1]\left[\ncodv\right] \nonumber\\
           &&\nonumber\\
           &\approx&~\frac{1.565\times 10^{-15}}{Z(\tex)}\,
 \frac{[{\rm e}^{(5.53\K)J/\tex} - 1]}
{{\rm e}^{(2.765\K)J(J+1)/\tex}} \nonumber \\
& & \left[\frac{\ncodv}{\cmtwo(\kms)^{-1}}\right].\label{tau_lte}
\end{eqnarray}
Here $\mu = 0.112\,$ debye is the electric dipole moment of the CO molecule
and $Z(\tex)$ is the CO rotational partition function:
\begin{equation}
Z(\tex) ~=~ \sum_{J=0}^{\infty} (2J+1)\,\exp[-(2.765\K)J(J+1)/\tex].
\label{partition}
\end{equation}

Using a grid of values of \ncodv\ and \tex, we have computed 
the vector {\bf r} of ratios under LTE conditions.  The series in Equation
\ref{partition} was truncated to 20 terms, since new terms added less than
1\% to the value of $Z$.  The \cor\ intensities were determined in the
same way as the \co\ intensities, except the \cor\ abundance was taken
to be 1/60 that of \co.  

\begin{figure*}[p!]
\plotone{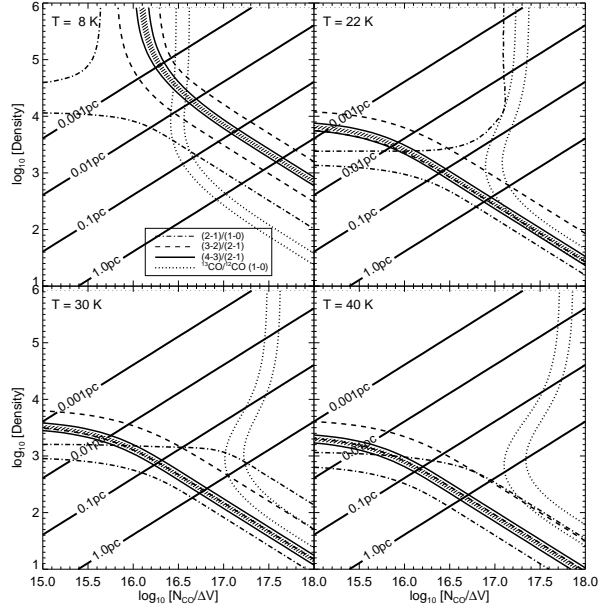}
\caption{Statistical equilibrium calculations for CO lines in clouds
with a large velocity gradient (LVG).  Each
plot maps out observed cloud CO line ratios as a function of column density
per unit velocity, $\nco/\Delta V$,
and gas volume density for a different temperature.  The shaded region 
indicates 
the range of values of the \co\ (4--3)/(2--1) ratio reported here for HLCs.  
 Straight lines indicate the total 
``size'' of the emitting regions, given by the ratio of column density 
to volume density, for CO-emitting cells with velocity width
$\Delta V = 1\kms$.  
The units of $\nco/\Delta V$ are \cmtwo\,$(\kms)^{-1}$.  Density
is in units of \cmthree.  
\label{cloud}}
\end{figure*}
\begin{figure*}[p!]
\plotone{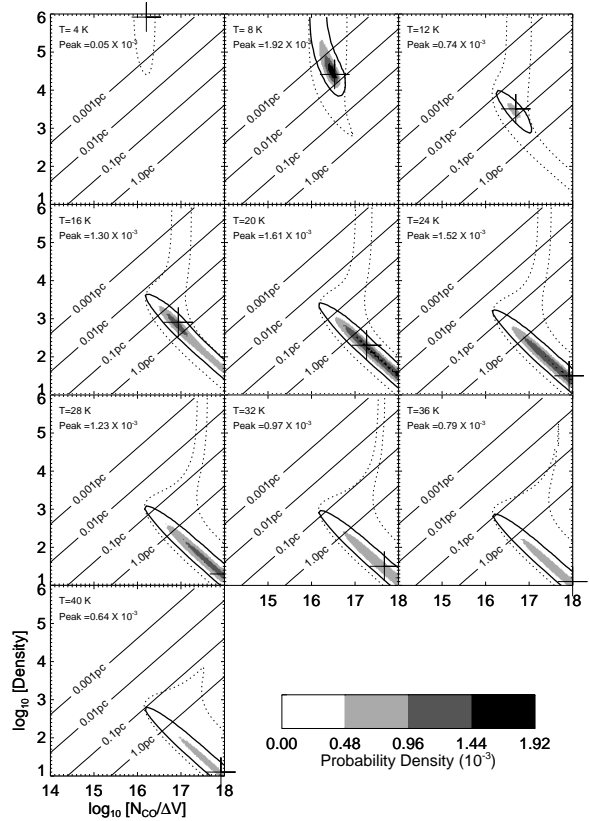}
\caption{The probability density field for HLC CO-emitting cells in the
LVG model.  Slices in the ($\nco/\Delta V$)--Density 
plane were made for 10 values of the temperature, and are represented as
greyscale images.  The dashed curves in each plot represent slices through
the 3$\sigma$ confidence surface for the sample region of 
[\tk,\ $\nco/\Delta V$,\ $n$]--space
when only HLC data for the (3--2) and lower transitions of \co\ and
\cor\ are taken into account.  The solid contours represent slices through
the 3$\sigma$ confidence surface when the new (4--3)/(2--1) data 
(this paper) are included.  The crosses indicate the point in each
slice where the probability density reaches its maximum value, listed
as ``Peak'' on each graph.  Straight lines indicate the ``size'' of the 
emitting regions, as described in Figure \ref{cloud}.
\label{cloud_models}}
\end{figure*}

{\plotfiddle{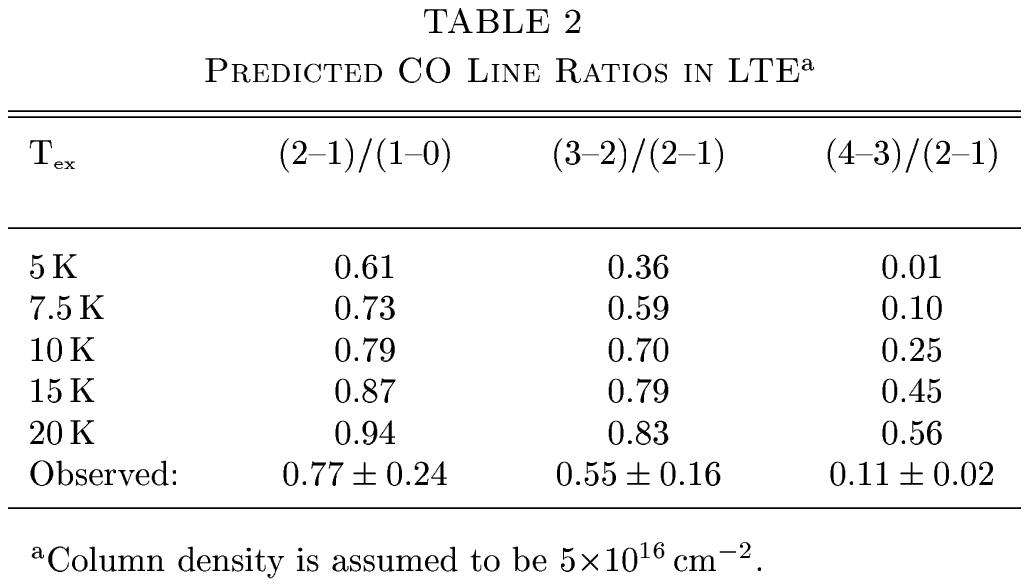}{2.5in}{0}{85}{85}{-135}{-235}}

Predicted \co\ LTE line ratios are listed in Table 2 
as a function of excitation temperature for $\nco = 5\times 
10^{16}\cmtwo$ and $\Delta V = 1\kms$.  The
bottom row of the table gives the actual values observed towards HLCs. 
The measured (2--1)/(1--0), (3--2)/(2--1), and (4--3)/(2--1) ratios are all 
compatible with low temperatures, $\tex\approx 5-10\K$.  Using the 
probability distribution function, 
$P({\bf r})d{\bf r}$ (Equation \ref{prob_dist}), we can determine
quantitatively a probability distribution in (\ncodv,\ \tex)--space.  
Figure \ref{cloud_lte} shows this.  The dashed contour on the
figure represents the 1$\sigma$ confidence interval (68.3\% of the
total probability) in the 
(\ncodv,\ \tex)--plane when only ratios 1, 2, and 4
are considered, \ie\ when we neglect the new (4--3) data (this paper).  
  The solid shaded contour represents the 1$\sigma$ confidence 
interval when the new data are included, 
and the two surrounding solid contours represent the 2 
and 3$\sigma$ confidence intervals (95.4\% and 99.7\%,
respectively, of the total probability).  The new data have 
decreased the size of the 1$\sigma$ contour
by a factor of $\sim 3$.  The most probable set of physical
properties is:  $\tex =  (8.5\pm 1.5)\K$ and 
$\ncodv = \expo{16.5\pm 0.2}\cmtwo\,(\kms)^{-1}$.
Evidently, if CO-emitting regions are in LTE then the temperature
is low, and the HDLT regime is the only solution.  The LTE line center 
opacity (Equation \ref{tau_lte}) is $\tau =$
9.6, 7.2, 2.1, and 0.3, for the \co\ (1--0), (2--1), (3--2), and (4--3)
transitions respectively.  As we deduced from the observations 
(see Figure \ref{cofits_b}), the CO-emitting regions are optically thick to
\co\ (1--0) radiation.  In fact, only the (4--3) and higher transitions of 
\co\ are optically thin.

\subsubsection{The LVG model}

The LTE analysis is necessary but not sufficient to 
confirm the HDLT scenario because the LTE model does not consider 
the possibility of low density gas.  We have simply shown that if the 
gas is dense enough such that the transitions are thermalized, then the 
temperature must be low.  There is no reason {\it a priori} to 
believe this is true.  If the temperature is high enough, intermediate 
levels such as $J=4$ may be significantly populated even in low density 
gas.  Furthermore, it may be radiative trapping which causes the 
thermalization of level populations, rather than high densities.  The LTE
model cannot distinguish between such possibilities.  

In this section we 
calculate the CO level populations in statistical equilibrium using the 
large velocity gradient (LVG) formalism (Sobolev 1960; Castor 1971).  This 
approximation states that the mean free path of spectral line photons, 
$l_{\litl mfp}$, is on average much less than the product 
$\Delta V \,(dV/dR)^{-1}$, where $\Delta V$
is the total cloud velocity dispersion and $dV/dR$ is the cloud
velocity gradient which is a systematic function of position.  In the 
LVG approximation different regions in a 
cloud are effectively decoupled in velocity space, whence line radiation
emitted from one region of the cloud does not interact with any other
region.  This simplifies considerably the radiative transfer computation.  

Of course, the LVG approximation does not strictly hold in HLCs---it is 
unlikely that the velocity field in HLCs is so well-ordered as to produce a 
systematic velocity gradient.  In general the clouds are not gravitationally 
bound (MBM; Keto \& Myers 1986; Pound \& Blitz 1993; Heithausen 1996),
nor do they possess evidence of large--scale outflows or rotation
(\eg\ Pound \etal\ 1991).  Instead, the 
velocity field in these clouds is most likely turbulent (Falgarone 1994; 
Falgarone \etal\ 1994), hence the velocity is a random function of position.  
Nevertheless, $l_{\litl mfp}$ is probably much 
less than the velocity correlation length of
the turbulence, \ie\ the spectral lines form under {\it macro-turbulent}
conditions [see for example the simulations of Park \& Hong 
(1995); see also the review in Falgarone \etal\ (1998)].  
Macro-turbulence prevents distinct emission regions of a cloud from 
``shadowing'' each other in velocity space, which
is statistically equivalent to a large systematic velocity gradient 
in its effect on the radiative transfer through the cloud.

Since the level populations both determine and are determined by the 
radiation field, the solution of the equation of radiative transfer for 
CO line radiation with self-consistent 
level populations was performed numerically.  The 
level populations in a statistical steady state, and the 
corresponding radiation field were computed for a 
grid of clouds, each with constant volume density,
$n$, kinetic temperature, $\tk$, and CO abundance, \xco.  At 
the surface the incident radiation field was taken to be that due to
the 2.7\K\ cosmic microwave background.  It was
furthermore assumed that there is no internal radiation source 
in the clouds.  Collisional rate coefficients for
CO excitation by \htwo\ were those computed by Flower \& Launay (1985).  A 
constant abundance $\xco = 8\times\expo{-5}$ was assumed, \ie\ 
in the CO-emitting regions approximately 80\% of the carbon gas is in CO.  
This is based on the local ISM gas-phase carbon abundance which is observed
to be [C]$=(1.4\pm 0.2)\times\expo{-4}$ (Sofia, Cardelli, Guerin, \& Meyer 
1997).  The results do not change drastically if \xco\ varies between 5 and 10
$\times\expo{-5}$.  The abundance of \cor\ was assumed to be 1/60 that of 
\co, \ie\ [\cor]/[\co]=[\carbon{13}]/[\carbon{12}].   We have ignored the
effects of chemical fractionation and isotope-selective 
photodissociation, which would cause the \cor/\co\ abundance
ratio to range between 1/70 to 1/35 (\S3.5; van Dishoeck \& Black 1988).

The populations of the first eight rotational transitions of the \co\ and
\cor\ molecules were modelled.  For each species the model was run for 
temperatures ranging from 4 to 40\,K, for
densities ranging from $10^1$ to $10^6$\cmthree, and for column density per
velocity ranging from $\ncodv = 10^{15}$ to $10^{18}$\cmtwo /(\kms).  
Results for $\tkin = 8$, 22, 30, and 40\K\ are displayed in Figure 
\ref{cloud}.  The observed range of values 
of the four ratios, $R_i \pm \sigma_i$ (defined in Equations 
\ref{R1} through \ref{R4}), are indicated by pairs of contours.  The range
in the (4--3)/(2--1) ratio is shaded.  For reference the size along
the line-of-sight, $L$, of CO emitting regions is 
indicated by straight solid lines.  A linewidth of $\Delta V = 
1\kms$ has been assumed, whence the following relationship gives the size:
\begin{equation}
\frac{L}{\pc}  \approx  4.05\times\expo{-2}
\left(\frac{\nco}{\expo{16}\cmtwo}\right)
\left(\frac{n}{\expo{3}\cmthree}\right)^{-1}
\left[\frac{\xco}{8\nexpo{-5}}\right]^{-1}.\label{size}
\end{equation}

Figure \ref{cloud_models} shows the statistical analysis of the LVG results.  
Using the method described, we have computed
the probability distribution function, $P({\bf r})d{\bf r}$, as a function
of \ncodv, $n$, and \tk.  The greyscale images in Figure \ref{cloud_models}
each represent a slice through the probability density 
field made in the (\ncodv,\ $n$)--plane at a particular temperature.  The 
dashed contour is a slice through the the 3$\sigma$ confidence 
surface if the (4--3)/(2--1) ratio is neglected and the solid contour 
represents a slice through the 3$\sigma$ confidence surface when the new 
data are included.  The crosses indicate the point in the 
(\ncodv,\ $n$)--plane where the probability density reaches its maximum 
value for the given temperature.  

\begin{figure*}[ht!]
\plotonesmall{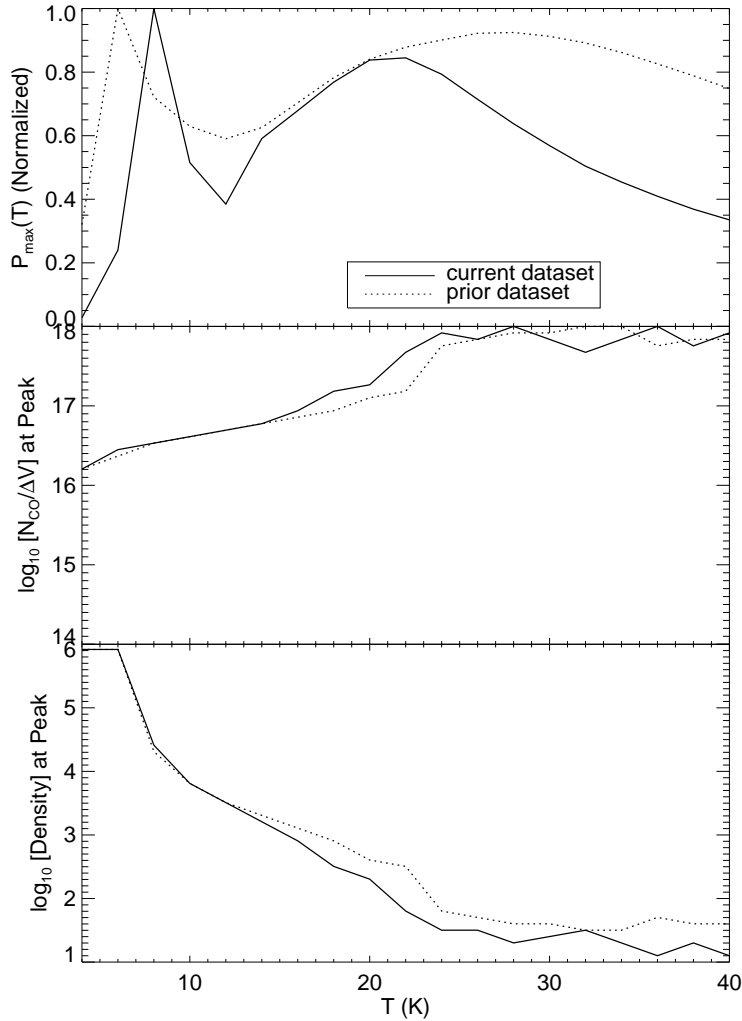}
\caption{Maximum-probability values of $P$, (\ncodv), and 
$n$ in slices of constant \tk\ (\ie\ the values of the quantities at 
the crosses shown in Figure \ref{cloud_models}).
Dashed curves represent these functions when only the (3--2) and lower 
transitions of CO are included, while solid curves include the new (4--3)
data.  The new data reduce the range of possible solutions,
but do not alter significantly the most probable cloud conditions.  Note
that $P_{max}(T)$ is normalized so that the peak value
equals unity.  The absolute peak probability density value is 4 times higher 
for the current dataset than for the prior dataset.
\label{cloud_prob}}
\end{figure*}

The maximum probability density in each temperature slice, 
$P_{\rm max}(\tk)$, \ie\ the value of the probability density at the
location of the crosses, is printed on the plots.  The probability
field has two local maxima, one at $\tk\approx 8\K$\
[$P_{\rm max}(\tk) = 1.92\times\expo{-3}$] and one at 
$\tk\approx 20\K\ [P_{\rm max}(\tk) = 1.61\times\expo{-3}]$.  These 
correspond to the HDLT and LDHT solutions, 
respectively.  The HDLT solution is the most probable.  This regime is
very well-defined in the (\ncodv,\ $n$)--plane, with
$n=\expo{4.5\pm 0.5}\cmthree$, and 
$\ncodv = \expo{16.6\pm 0.3}\cmtwo(\kms)^{-1}$.  It is reassuring that 
we get the same value of \ncodv\ in the HDLT/LVG model
as in the LTE model.  The two models tell
the same story:  if the density is high then the level populations are 
thermalized in constant-opacity structures with $\tk\approx 8\K$.  The 
\co\ line center opacities in the LVG model are:  
 $\tau=$ 9.6, 14.4, 6.3, and 1.0 for the (1--0), (2--1), (3--2), and (4--3) 
transitions, respectively.  In this model all \co\ radiation is produced in 
gas which is at least marginally optically thick.

 As mentioned by FP the implied structure sizes in the HDLT case are 
very small, less than $\sim 0.01\pc$ ($\sim 2000\,$AU), which is consistent
with the $< 6500\,$AU--sized structures seen by Pound \etal\ (1991)
in high--latitude cloud MBM-12.  This HDLT structure size is
$\lesssim$1\% of the average projected CO size of HLCs ($\sim 1\pc$; 
Ingalls 1999).  Thus in this interpretation a typical HLC is comprised of
an ensemble of $\sim 10,000$ HDLT CO-emitting cells (given an area
filling fraction of unity).  

Unfortunately, it is still not possible to rule out the LDHT solution
completely, since the probability field has a 
second local maximum at $\tk\approx 20\K$ which is about 80\% as probable
as the 8\K\ maximum.  We are, however, able to place a strict upper limit on 
the temperature in the CO-emitting regions due to the predicted size
of LDHT structures.  As mentioned in \S1, the sizes of 
emitting structures along the line-of-sight (LOS) may be much larger
than their sizes in the plane of the sky.  It is not plausible, however, for
LOS structure sizes to exceed the $\sim 1\pc$ {\it cloud} sizes.  
  For very high temperatures ($\tk\gtrsim 30\K$) the LOS size of emitting 
regions with nonzero probability in the model becomes unrealistically 
large.  Unless most observed CO structures are filaments aligned along 
the line of sight, which is unlikely, temperatures greater than $\sim 30\K$ 
must be completely ruled out.  

Again (see the LTE model) the \co\ (4--3) data significantly decrease the 
size of the 3$\sigma$ surface in the parameter space.  In other words, 
the {\it range} of possible solutions is decreased.  Remarkably, 
including the higher transition data does not change the 
predicted physical conditions.  We plot in Figure 
\ref{cloud_prob} the maximum-probability values of $P$, (\ncodv), and 
$n$ (the values at the positions of the crosses).  As for Figures
\ref{cloud_lte} through \ref{cloud_models}, dashed
curves represent these functions when only the (3--2) and lower transitions
of CO are included, and solid curves represent the same functions when the
new (4--3) data are used.  Clearly the new data both limit the range 
of possible solutions in \tk -space (Figure \ref{cloud_prob}, {\it top}) 
and also yield cloud conditions that are entirely 
consistent with those derived using the old data (Figure \ref{cloud_prob},
{\it middle} and {\it bottom}).

\section{Discussion}

\subsection{Physical Implications of HDLT Gas}

We have studied the J = (1--0) through (4--3) transitions of 
CO in translucent clouds.  Motivated by the work 
of Falgarone \etal\ (1991; 1992; 1998) and FP, instead of treating 
clouds as separate entities with distinct characteristics, we sought 
regularity in the CO emission.  The linear relationships among the 
(1--0), (2--1), (3--2), and (4--3) transitions, as well as the 
constancy of the 60\micron/100\micron\ {\sl IRAS} surface brightness 
ratio, justify
this perspective.  The simplest interpretation of these results is that
the emission arises in regions with the same physical properties, or
``cells'', and that variations in the emission are caused by variations 
in the beam filling fraction of the cells.  We call this the 
``cell hypothesis''.  

Two types of CO cell may exist:  the low density, high temperature (LDHT)
and the high density low temperature (HDLT) cell.  The LDHT cell is 
relatively space-filling, with line of sight sizes of order 50\% to 100\% 
of the cloud size.  It is also in thermal pressure equilibrium with the 
interstellar environment.

The existence of HDLT cells, on the other hand, is controversial.  We have
shown that cells with densities of order $\expo{4.5}\cmthree$
are the most probable source of CO emission.  This challenges the picture of 
HLCs described by van Dishoeck \etal\ 
(1991), who argued against the possibility of gas with 
$n \sim \expo{4}\cmthree$.  The HDLT scenario also
defies one of the primary assumptions of interstellar physics, which
holds that clouds are in hydrostatic equilibrium.  HDLT cells have
a thermal pressure $P/k\sim 2\nexpo{5}\cmthree\K$, which is an
order of magnitude greater than the ambient thermal ISM pressure.  Is this
a serious problem?

The interstellar medium is turbulent.  Its structures are not
the result of equilibrium processes.  Recent simulations of 
magnetohydrodynamic (MHD) turbulence performed by
Ballesteros-Paredes, V\'asquez-Semadeni, \& Scalo (1999) show that
thermal pressure equilibrium is irrelevant, since in MHD turbulence
the volume and surface
terms of the Eulerian Virial Theorem are comparable.  As a consequence, structures
will nearly always deform under inertial motions, resulting in ``clumping''
on many scales.  According to this model, in a turbulent cloud, clumps are continually
forming due to density fluctuations and dissipating due to radiative 
cooling.  Quasi-static 
configurations are highly unlikely, unless compressed to sizes 
$L\lesssim 0.01\pc$, where self-gravity becomes important.  Ironically, the 
HDLT cells are approximately this size.  If the simulations of 
Ballesteros-Paredes \etal\ (1999) are correct, then {\it the only stable 
configuration in translucent molecular clouds is the HDLT one}.  This
important result needs to be confirmed.  Is the CO emission produced 
only in stable structures which are in LTE?  This would explain the 
remarkable uniformity we see in the 
CO line ratios for a large sample of clouds.

\subsection{Limitations of the Method}

We have attempted in this paper to evaluate two competing
models for the physical state of molecular gas in translucent clouds,
the HDLT and LDHT models.  Both van Dishoeck \etal\ (1991) and 
FP state that neither HDLT nor LDHT conditions uniquely describe the data
(even though each group concludes by favoring a different configuration;
see \S1).  We have assumed that clouds are macro-turbulent, 
and we have argued that the LVG radiative transfer model is an adequate 
approximation of macro-turbulence.  By deriving a probability density 
field in the physical parameter space we have shown quantitatively that
HDLT gas is the more likely source of observed CO emission.

Our analysis is limited, however, by the extent to which the LVG scenario 
is an accurate representation of the velocity field of translucent gas.  The 
assumption that clouds are macro-turbulent
is based on the extensive three-dimensional radiative transfer simulations 
done by Park \& Hong (1995).  Nevertheless, it can be argued that using the 
macro-turbulent LVG assumption has biased our results in favor of HDLT 
conditions, since tiny HDLT cells are more likely to satisfy
the LVG criterion, because they
are less likely to shadow each other along the line of sight than the 
much larger ($\sim 0.5-1\,$pc) LDHT cells.  Furthermore, since they
have larger values of $\nco/\Delta V$, LDHT cells have
a much larger column density of radiatively coupled gas than HDLT cells 
with the same internal velocity dispersion.  Thus the macro-turbulent 
approximation should break down for the largest LDHT cells.  
A more complete study would vary the statistics of the velocity field
between the micro-turbulent and macro-turbulent limits.  This is 
beyond the scope of the current paper, and we leave it to future work.

\subsection{Testing for HDLT Cells}

If HDLT cells exist, then they can be detected directly in CO
emission using a millimeter-wave interferometer.  A {\it resolved} HDLT cell
would have a Rayleigh-Jeans brightness temperature in \co\ (1--0)
of $\ta\approx 5\K$.  An interferometer with
a spatial resolution of 0.001\pc\ (for a 
cloud at a distance of 100\pc\ this corresponds to an angular resolution 
of 2\arcsec ) would be able to detect such HDLT cells.  
If there is no detectable \co\ (1--0) emission in HLCs at 2\arcsec\ scales, 
then the HDLT hypothesis is false.  

\acknowledgments
This work forms part of the PhD Thesis of JI.  We thank Dan Clemens, 
James Jackson, and Harlan Spence for critical reading of an
early version of the manuscript.  We thank William Reach for helpful 
discussions.  We thank the referee, Barry Turner, for careful reading
of the manuscript, and for suggestions which have improved the final
product.  We thank Mario Tafalla for his LVG code.  We also thank 
the 1997 winter crew of Amundsen-Scott South 
Pole Station.  This research was supported in part by the National Science 
Foundation under a cooperative agreement with the Center for Astrophysical 
Research in Antarctica (CARA), grant number NSF DPP 89-20223.  CARA is a 
National Science Foundation Science and Technology Center.

\end{document}